\documentclass[aip,jcp,reprint]{revtex4-1}
\usepackage{graphicx} 
\usepackage{natbib}
\usepackage[utf8]{inputenc}
\usepackage[english]{babel}
\usepackage{amsmath}
\usepackage{amsfonts}
\usepackage{amssymb}
\usepackage[T1]{fontenc}
\usepackage{mathrsfs}
\usepackage{braket}
\usepackage{geometry}
\usepackage{here}
\usepackage{lmodern}
\usepackage{array}
\usepackage{url}
\usepackage{textcase}
\usepackage{bm}
\usepackage{xcolor}

\usepackage{xcolor}

\usepackage{graphicx} 
\usepackage{multirow}
\usepackage{booktabs}

\usepackage{graphicx} 
\newcommand{\modif}[1]{\textcolor{black}{#1}}
\begin{document}
\author{Brieuc Le D\'e}
 \affiliation{Institut des Nanosciences de Paris, Sorbonne Université, CNRS, F-75005 Paris, France};
 \affiliation{Department of Chemistry and Biochemistry, University of California, Merced CA 95343, USA}

\author{Etienne Mangaud}
 \affiliation{MSME, Universit\'e Gustave Eiffel, UPEC, CNRS, F-77454 Marne-La-Vall\'ee, France}

\author{Alex W. Chin}
\affiliation{Institut des Nanosciences de Paris, Sorbonne Université, CNRS, F-75005 Paris, France}

\author{Mich\`ele Desouter-Lecomte}
\email{michele.desouter-lecomte@universite-paris-saclay.fr}
\affiliation{Institut de Chimie Physique, Universit\'e Paris-Saclay-CNRS, UMR8000, F-91400 Orsay, France}

\title{Revisiting crossed-correlated baths in open quantum systems simulated by HEOM or T-TEDOPA}

\begin{abstract}
Excited-state dynamics of open quantum systems is analyzed by the hierarchical equations of motion (HEOM) or the thermalized time-evolving density operator with orthogonal polynomials algorithm (T-TEDOPA) method when a discrete \textit{ab initio} linear vibronic model is parametrized by continuous temperature-dependent spectral densities leading to crossed correlation functions, i.e. correlated fluctuations of the energy gap collective modes. We focus on a conical intersection involving two collective modes tuning the energy of each excited state and we revisit the  transformation of the initial correlated tuning baths to de-correlated shared baths in order to reduce the computational resources. While a completely frequency-dependent transformation poses problems for HEOM, we find that in some particular cases, an optimal approximate frequency-independent transformation may be derived. On the contrary, T-TEDOPA is very efficient and allows to use this frequency-dependent transformation at the price of managing long-range couplings in the tensor chain.  An illustrative application is shown by using the linear vibronic coupling model of a planar symmetrical  (phenylethynyl)benzene dimer.      
\end{abstract}

\maketitle

\section{Introduction}
Simulating dynamics of complex systems has been and remains a computational challenge that has given rise to a flowering of algorithms and software. Since measured observables often depend on a limited number of degrees of freedom (DoF), a customary starting point is the partition of the complete system into a main subsystem that gathers the active DoFs and an environment, that is, the system-bath model \cite{Breuer2002,Weiss2012,Kuhn2011}.  From the quantum mechanical method, the first step is the definition of a model Hamiltonian symbolically separated in three parts related to the relevant subsystem, the reservoir, and the system-bath interaction. Since the work of Caldeira-Leggett \cite{Leggett1987}, the reservoir is usually an ensemble of harmonic oscillators, and the interaction term is bilinear in the system and the bath operators. \modif{In this pioneering model, two electronic states (denoted spin, i.e. spin $1/2$) are coupled to a harmonic reaction coordinate explicitly included in the system Hamiltonian and  coupled to a set of secondary bath oscillators. A great alternative already discussed in a seminal paper on electron transfer \cite{Garg1985} restricts the system to the two-electronic states at a given reference nuclear geometry and coupled to a new set of non interacting oscillators obtained by a unitary transformation. This generates the so-called « star model » with all the oscillators (phonon modes or all the molecular and solvent vibrational modes) coupled to the electronic subspace, while the first model is generally referred to as the reaction coordinate mapping method \cite{Nazir2014,Nazir2016,Chenel2014,Eisfeld2015,Gelin2016,Ikeda2022}. The star model has been used to simulate molecular excited state dynamics for instance excitonic states in light-harvesting molecular structures \cite{Fleming2010}, quantum dots \cite{Lovett2021}, or excited states in chromophores coupled by non-adiabatic interstate interactions \cite{Thorwart2016,MDL2019,Gao2024}.} 

\modif{We focus here on the dynamics of electronic excitation transfer via a conical intersection, which leads to strong non-adiabatic coupling \cite{Domcke2011,Baer2006}. When the dynamics is treated by the wave function approach, the linear vibronic coupling (LVC) model is the basic tool for many simulations \cite{Picconi2024,Tremblay2022,Burghardt2025,Galiana2023,Galiana2024,Galiana2025,Padmabati2025} with Multi Layer Multi Configuration Time Dependent Hartree (ML-MCTDH) \cite{Meyer1990} or Matrix Product State (MPS) computational methods \cite{Liu2024,Gao2024,DunnettMB2021,Zuehlsdorff2024} involving some hundreds of vibrational DoFs. In the open quantum system simulations with the hierarchical equations of motion (HEOM) \cite{Kubo1989,Ishizaki_2005,Yan2007,Yan2009,Tanimura2006,Tanimura2020,Shi2024}, the simulation of dynamics via a conical interaction is challenging since at least three baths are involved: two baths make the energy of the electronic states fluctuate and a third bath strongly alters the electronic coupling.  A discussion between the two alternative models including or not active modes in the open system was presented in the reference \cite{Thorwart2016}. The partition retaining a two-dimensional subspace of nuclear degrees of freedom where the degeneracy of the conical intersection is lifted coupled to a residual bath is adopted in ref. \cite{Gelin2016}. The star model where all the nuclear degrees of freedom are in the baths leads to stronger system-bath couplings and requires more computational resources \cite{Thorwart2016,MDL2019}. } 

\modif{When the environmental effects are taken into account, the central tool is the spectral density matrix, whose diagonal terms indicate the strength of the system-bath coupling at each bath frequency and off-diagonal terms are related to the correlation of the effects of a mode at a given frequency on two different states. Indeed, the intramolecular modes often show strong correlations, in the sense that displacement of these modes either reduces or increases the energy gap between distinct electronic states. The spectral density matrix may be constructed from discrete $\textit{ab}$ $\textit{initio}$ data, broadened to generate a continuous distribution and possibly sampled again to generate fewer modes than in the initial model for wave packets simulations \cite{Burghardt2010,Burghardt2019}. An alternative is to directly obtain a continuous matrix by transforming correlation functions calculated by molecular dynamics, and then possibly proceed with sampling \cite{Firmino2016,Mendive2018,Geva2020,DunnettMB2021,Zuehlsdorff2024}. This approach allows the explicit inclusion of solvent effects, vibration damping, and even conformational changes to be accounted for via the effective (continuous) spectral function matrix that contains information about the energy fluctuations arising from all environmental influences. }
 
 Bath correlation is ubiquitous in the modelling of condensed-matter systems and has been considered mainly in the context of excitons in multi-site Hamiltonian simulating light-harvesting systems or biological environment \cite{Ishizaki2010,Nalbach2010,Silbey2010,Silbey2012,Schulten2011,Mukamel2011,Fleming2010,Egorova2012,Plenio2017,Makri2023}, in qubit physics \cite{Ting2015,Burkhard2025} and also condensed matter, i.e. electron-phonon interaction in solids/metals/semiconductors - the basis of pairing in superconductivity. These strong (spatial) correlations in bath fluctuations naturally arise from the extended (delocalized, plane wave) nature of lattice phonons \cite{Lacroix2021}, but can be still simulated in thermalized time-evolving density operator with orthogonal polynomials algorithm (TEDOPA) by including long-range interactions \cite{LacroixLovett2024}.
  The problem arises both in sample-based wave packet simulations by MCTDH  \cite{Domcke2011,Burghardt2019} or MPS \cite{DunnettMB2021,Zuehlsdorff2024} and in dissipative dynamics with memory kernels treated by HEOM \cite{Kubo1989,Ishizaki_2005,Yan2007,Yan2009,Tanimura2006,Tanimura2020,Shi2024,Yan2002,Yan2005,SongShi2017,Tanimura2024}.   
 
\modif{In this work, we address the central issue of bath de-correlation procedure by HEOM and T-TEDOPA formalisms. A first idea could be to propose a transformation of the collective bath modes to diagonalize the correlation matrix or the associated spectral density matrix. However, whether the transformation used to diagonalize the spectral density matrix is frequency dependent or frequency independent is a crucial point within the HEOM framework. We shall mention situations where this diagonalization is possible and propose different approximate strategies useful to reduce the computational cost by involving a simple set of independent environments compatible with the standard HEOM formulation. Interestingly, we will show that this frequency dependence could be taken into account in T-TEDOPA. We will compare this procedure directly based on the frequency dependent diagonalization with a recent alternative method using T-TEDOPA in a simulation of a conical intersection\cite{Tamascelli2019,Dunnett2021,Lacroix2024}. The goal is then to extract parameters of a system-bath coupling terms from the spectral density matrix calculated by molecular dynamics. This strategy leads to an operator that explicitly couples a given electronic state to different baths as it is the case for spatial correlations. This illustrate the flexibility to write the system-bath coupling with transformed modes and operators. This will be the heart for the de-correlation procedure in the HEOM or T-TEDOPA formalisms that have different numerical constraints.  }    

As illustration, we reconsider the LVC model of a conical intersection (sec.\ref{sec:LVCmodel}) in a symmetric meta-substituted planar poly(phenylene-ethylene) (PPE) dimer already used in different applications with MCTDH \cite{Galiana2023} or HEOM \cite{Jaouadi2022,LeDe2024}. The dynamics \modif{of} two excited states depend on three baths: the two tuning baths that \modif{make the energy gap with the ground state fluctuate} and the coupling bath, which causes the variation of the electronic interstate coupling.  In recent HEOM simulations \cite{Jaouadi2022,LeDe2024}, the two correlated tuning baths were replaced by a single shared bath with a modified system operator by an approximate frequency independent transformation of the baths. Here, we intend to justify this approach. 

We first simulate the excited state dynamics by HEOM (secs.\ref{sec:crosscorrelation} and \ref{sec:HEOM}) including all the crossed terms which is rarely done in HEOM simulations \cite{Yan2002,Yan2005} where approximate treatments are used \cite{Ishizaki2010,Fleming2010,Schulten2011,Jaouadi2022,LeDe2024}. Then, we justify and discuss the validity of the approximate strategy, which is very useful to reduce the computational resources in HEOM simulations (sec.\ref{sec:Transfo}). The efficiency of the T-TEDOPA method to apply the exact frequency dependent transformation leading to a single shared bath in favorable cases is presented in sec.\ref{sec:T-TEDOPA}. This strategy in T-TEDOPA is then an alternative to both the exact correlated HEOM method and to another approximate de-correlation procedure recently proposed \cite{DunnettMB2021,Hunter2024_ie,Zuehlsdorff2024}. 

\section{Fully (anti) correlated baths}
\label{sec:fully/fullyanti}
\begin{figure*}
 \centering
\includegraphics[width =0.8\textwidth,height=7cm]
{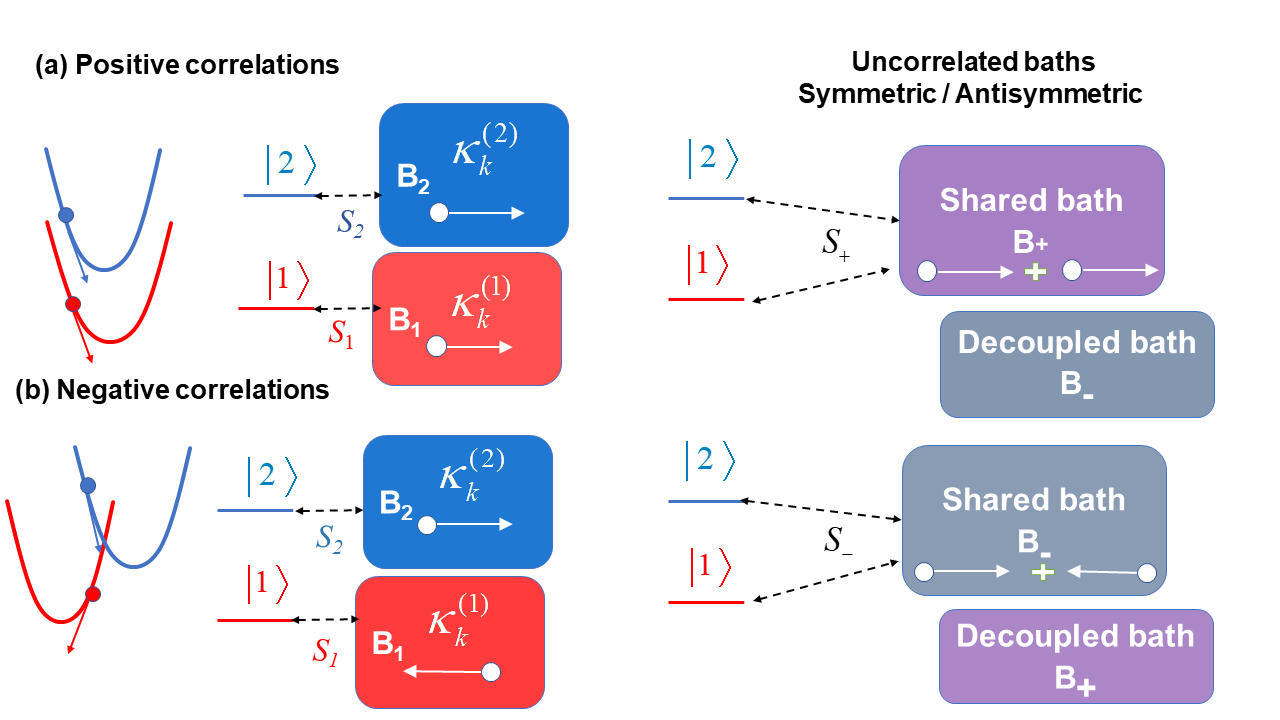}
\caption{Collective bath modes transformation for fully correlated or fully anti-correlated baths. (a) Positive correlations, the bath modes are in phase. the shared bath is the symmetric combination $B_+$. (b) Negative correlations, the modes have opposite phases and the shared bath is the antisymmetric combination $B_-$. The positive or negative correlations are illustrated in the framework of a LVC model.  }  
\label{fig:modeltwobaths}
\end{figure*}

We start with a simple example of a two-level system \modif{described by the LVC model} in which each state $\left| n \right\rangle $ is coupled to a bath of oscillators as schematized in fig.\ref{fig:modeltwobaths}. The partition of the full model leads to the generic Hamiltonian
\begin{equation}
H={{H}_{S}}+{{H}_{B}}+{{H}_{SB}}
\label{eq:Htotal}
\end{equation}
where ${{H}_{S}}=\sum\nolimits_{n=1}^{2}\epsilon_n{\left| n \right\rangle }\left\langle  n \right|$ is the system Hamiltonian, ${{H}_{B}}$ is the environment Hamiltonian, i.e., the sum of the harmonic baths ${{h}_{n}}=\frac{1}{2}\sum\nolimits_{j}{\left( p_{n,j}^{2}+\omega _{n,j}^{2}q_{n,j}^{2} \right)}$ expressed in mass weighted coordinates. \modif{in the LVC model, the $q_j$ are normal modes assumed to be the same in each electronic state}. By \modif{discarding any electronic interstate coupling, the two tuning baths} are coupled only diagonally to the system, the system-bath coupling \modif{may always take the form of a sum of bilinear terms}
\begin{equation}
{{H}_{SB}}=\sum\nolimits_{n=1}^{2}{{{S}_{n}}}{{B}_{n}}	
\label{HSBnew}
\end{equation}
where the system-bath coupling operators are projectors on the basis states ${{S}_{n}}=\left| n \right\rangle \left\langle  n \right|$. The collective bath modes are linear combinations of the bath coordinates 
\begin{equation}
B_{n}=\sum\nolimits_{j}{\kappa_{j}^{(n)}q_{j}}
\label{eq:Collectivebathmode}
\end{equation}
where $\kappa _{j}^{(n)}$ is the gradient \modif{making the energy of the corresponding state fluctuate}. 

\modif{The fact that $H_{SB}$ is a simple sum of terms does not mean that the baths are not correlated. Indeed, even if the normal modes $q_k$ are independent harmonic oscillators, the operators $B_i$ and $B_j$ can still be correlated whenever they share common modes. The two baths emerging from the LVC Hamiltonian are naturally correlated: they are not distinct physical environments, but different projections of the same vibrational environment onto different electronic states.} 

The Gaussian statistical properties of the harmonic baths are characterized by the two-time correlation matrix
\begin{equation}
{{C}_{lm}}(t-\tau )=\left\langle {{B}_{l}}(t){{B}_{m}}(\tau ) \right\rangle 
\label{eq:C(t)new}
\end{equation}
where $l,m\in \left[\!\left[ 1,2 \right]\!\right]$ and $\left\langle \centerdot  \right\rangle $ is the average over the equilibrated baths at temperature $T$ described by  
\begin{equation}
\rho _{B}^{eq}(T)=\exp \left( -\beta {{H}_{B}} \right)/T{{r}_{B}}\left[ \exp \left( -\beta {{H}_{B}} \right) \right]
\label{eq:rhoeq}
\end{equation}
with $\beta =1/{{k}_{B}}T$ and ${{k}_{B}}$ is the Boltzmann constant.

In the frequency space, the correlation matrix is linked to the temperature-dependent spectral density
\begin{equation}
{{C}_{lm}}(t-\tau )=\frac{\hbar }{\pi }\int_{-\infty }^{\infty }{d\omega J_{lm}^{\beta }(\omega )}{{e}^{-i\omega (t-\tau )}}	
\label{eq:TFC(t)new}
\end{equation}
with
\begin{equation}
J_{lm}^{\beta }(\omega )={{J}_{lm}}(\omega ){{n}_{\beta }}(\omega ) 
\label{eq:Jbeta}
\end{equation}
where ${{n}_{\beta }}(\omega )$ is the Bose function and $J_{lm}^{{}}(\omega )$ is the temperature-independent spectral density \modif{that is first defined in a discrete representation}:
\begin{equation}
{{J}_{lm}}(\omega )=\frac{\pi }{2}\sum\nolimits_{j}{{g_{j}^{(l)}g_{j}^{(m)}}}\,\delta (\omega -{{\omega }_{j}})
\label{eq:spectraldensity}
\end{equation}    
where $g_{j}^{(l)}$  are the system-bath coupling given by $\kappa _{j}^{(l)}\omega_j^{-1/2}$ for $l=1,2$. \modif{The spectral density matrix contains information about the energy fluctuations arising from all environmental influences, including intramolecular or solvents modes. Intramolecular modes often show strong correlations, although some fluctuations affect the electronic states in effectively random (uncorrelated) ways (especially true of the solvent contributions). When constructing the spectral density matrix from the LVC model coefficients, the off-diagonal element $J_{12}(\omega)$ becomes non-zero because the two electronic states couple to overlapping subsets of vibrational coordinates.  However, when the spectral densities are smoothed by a broadening procedure, the continuous off-diagonal terms may become small due to cancellations between positive and negative mode-specific contributions. As a result, the effective bath correlation perceived by the dynamical model can become significantly weaker than the underlying LVC model. }

\modif{The interesting inverse procedure followed in refs.\cite{DunnettMB2021,Zuehlsdorff2024} starts from the spectral density matrix known from molecular dynamics calculations, including both the diagonal terms and the cross term $J_{12}(\omega)$. The goal is then to reconstruct an explicit microscopic bath Hamiltonian that reproduces these correlations. To do this, one introduces a set of harmonic modes and assigns coupling coefficients such that each mode can couple simultaneously to several system operators. The $H_{SB}$ operator is then similar to the one used in non-local correlations when a given electronic state is coupled to oscillators on different sites. \cite{Fleming2010,Ishizaki2010}. In practice, this means finding a set of couplings whose  reproduce the spectral density matrix. Conceptually, this is a factorization problem that may be realized by different ways. The shared modes generated by this factorization are what produce the correlated baths. }

One may consider two extreme situations where the baths are fully correlated, this means the collective modes are always in phase or fully anti correlated if they have opposite phases. The gradients are equal in each bath and have the same sign in the first case or opposite signs in the second one (see fig.\ref{fig:modeltwobaths}). Assuming fully correlated or fully anti-correlated baths means $g_{j}^{(1)}=\pm g_{j}^{(2)}\ \forall j$. The ${{J}_{lm}}(\omega )$ matrix is then of rank 1 for each frequency and it may be diagonalized  by a single frequency-independent matrix
\begin{equation}
 {{\mathbf{J}}^{\beta }}(\omega )=\mathbf{U}\mathbf{\Lambda} (\omega ){{\mathbf{U}^T}}
 \label{eq:DiagJbeta}
\end{equation}
where $\mathbf{\Lambda (\omega )}$ is the diagonal matrix of the eigenvalues, which contains only one non-zero eigenvalue in this case. By this orthogonal transformation one may build new collective bath operators ${{\tilde{B}}_{j}}(t)=\sum\limits_{n}{U_{nj}}{{B}_{n}}(t)$ that remain uncorrelated at all times. Moreover, only the new bath associated with the non-zero eigenvalue is coupled to the system. The two initial baths are compressed in a single active shared bath, the other one being decoupled from the system. In the fully correlated or fully anti-correlated case, the shared bath is 
\begin{equation}
 {{{B}}_{\pm }}(t)=\frac{1}{\sqrt{2}}\left( {{B}_{1}}(t)\pm {{B}_{2}}(t) \right)
 \label{eq:B+-}
\end{equation}
with the $+$ or $-$ sign respectively. The $H_{SB}=\textbf{s}^T\textbf{b}$ operator (with $\textbf{s}$ and $\textbf{b}$ the column vectors of the operators and of the collective bath modes) may be written $H_{SB}= \textbf{s}^T\mathbf{U}\mathbf{U}^T\textbf{b}$. The transformed system operator is then
\begin{equation}
{{{S}}_{\pm }}=\frac{1}{\sqrt{2}}\left( {{S}_{1}}\pm {{S}_{2}} \right)
\label{eq:Stilde}
\end{equation}
(see fig.\ref{fig:modeltwobaths}). Note that the standard spin boson model generally uses a single shared bath coupled to the system by the ${{\sigma }_{z}}$ Pauli matrix, which corresponds to a fully anti-correlated case.

\section{Cross-correlation in master equation }
\label{sec:crosscorrelation}
We now consider a generic d-level system coupling to $N_{bath}$ harmonic baths
\begin{equation}
{{H}_{SB}}=\sum\nolimits_{\alpha }^{{{N}_{bath}}}{{{S}_{\alpha }}}{{B}_{\alpha }}. 
\label{eq:HSB}
\end{equation}
The subsystem is described by the reduced density matrix ${{\rho }_{S}}(t)=T{{r}_{B}}\left[ \rho (t) \right]$ obtained by tracing the bath DoFs.  In the interaction representation, the evolution of the system density matrix ${{\tilde{\rho }}_{S}}(t)$ is driven by a dynamical map, which involves a positive time ordered exponential denoted by ${{\mathcal{T}}_{+}}$ 
\begin{equation}
{{\tilde{\rho }}_{S}}(t)={{\mathcal{T}}_{+}}{{e}^{\int\limits_{0}^{t}{d\tau \left\langle L(\tau ) \right\rangle }}}{{\tilde{\rho }}_{S}}(0) 
\label{eq:rhoS}
\end{equation}
where $\left\langle \centerdot  \right\rangle $ means trace over the bath DoFs. In this representation, one has 	$L(t)\centerdot =\frac{-i}{\hbar }\left[ {{H}_{SB}}(t),\centerdot  \right]$ with 	${{H}_{SB}}(t)=\sum\nolimits_{j}^{{{N}_{bath}}}{{{S}_{j}}(t)}{{B}_{j}}(t)$, 
	${{S}_{j}}(t)={{e}^{i{{H}_{S}}t}}{{S}_{j}}{{e}^{-i{{H}_{S}}t}}$ 
and ${{B}_{j}}(t)={{e}^{i{{h}_jt}}}{{B}_{j}}{{e}^{-i{{h}_jt}}}$.  
According to the Wick theorem, the cumulant expansion of Eq.(\ref{eq:rhoS}) is exactly truncated at the second order if the statistics is Gaussian as it is the case for harmonic baths. Therefore, Eq.(\ref{eq:rhoS}) becomes
\begin{equation}
{{\tilde{\rho }}_{S}}(t)={{\mathcal{T}}_{+}}{{e}^{{{F}^{(2)}}(t,{{t}_{0}})}}{{\tilde{\rho }}_{S}}(0)
\label{eq:rhocumul}
\end{equation}
with the second order cumulant 
\begin{equation}
 {{F}^{(2)}}(t,{{t}_{0}})=\int_{{{t}_{0}}}^{t}{d\tau }\int_{{{t}_{0}}}^{\tau }{d\sigma }{{K}^{(2)}}(\tau ,\sigma )   
\end{equation}
where \modif{the memory kernel is defined} involving only two-time correlation functions  
\begin{align}
  & {{K}^{(2)}}(t,\tau )\centerdot =T{{r}_{B}}[L(t)L(\tau )\rho _{B}^{eq}\centerdot ] \nonumber \\ 
 & =-\frac{1}{{{\hbar }^{2}}}\left\langle \left[ {{H}_{SB}}(t),\left[ {{H}_{SB}}(\tau ),\centerdot  \right] \right] \right\rangle  
 \label{eq:kernel}
\end{align}
and the trace is made over each bath. The master equation derived from the mapping of Eq.(\ref{eq:rhocumul}) reads:
\begin{equation}
{{\dot{\tilde{\rho }}}_{S}}(t)={{\mathcal{T}}_{+}}\int\limits_{0}^{t}{d}\tau {{K}^{(2)}}(t,\tau ){{\tilde{\rho }}_{S}}(t).    
\end{equation}
By inserting Eq.(\ref{eq:kernel}) in Eq.(\ref{eq:rhocumul}), one gets :
\begin{align}
  & {{\hbar }^{2}}{{{\dot{\tilde{\rho }}}}_{S}}(t) \\ 
 & =-\sum\limits_{l,m=1}^{{{N}_{bath}}}{\left[ {{S}_{l}}(t), \right.}\int\limits_{0}^{t}{d}\tau \left( {{C}_{lm}}(t,\tau ){{S}_{m}}(\tau ){{{\tilde{\rho }}}_{S}}(t) \right. \nonumber \\ 
 & \left. -{{C}_{ml}}(\tau ,t){{{\tilde{\rho }}}_{S}}(t){{S}_{m}}\left. (\tau ) \right) \right].
 \label{eq:mastereq}
\end{align}
Even if the system-bath coupling ${{H}_{SB}}$ (Eq.(\ref{eq:HSB})) is the sum of bath contributions that seem independent, the master equation (Eq.(\ref{eq:mastereq})) rigorously involves the full correlation matrix of the bath collective modes ${{C}_{lm}}(t,\tau )$ (Eq.(\ref{eq:C(t)new})). 
The statistical properties are well known \cite{Yan2002,Yan2005} (they are summarized in the \modif{supplementary material}). For two particular modes $m,l$ the main relations are :
\begin{equation}
{{C}_{ml}}(\tau ,t)={{C}_{lm}}(t-i\hbar\beta,\tau  )=C_{lm}^{*}(t,\tau ).    
\end{equation}


\section{HEOM for correlated baths }
\label{sec:HEOM}
The crucial point of the HEOM method is the expansion of the correlation functions in terms of oscillatory decaying complex exponential functions \cite{Yan2009,Shi2014,Yan2022,ThossBorrelli2024,Tokieda2025}, each term being associated with an artificial decaying mode
\begin{equation}
{{C}_{lm}}(t,\tau )=\sum\nolimits_{k}^{{{K}^{(lm)}}}{\alpha _{k}^{(lm)}}{{e}^{i\gamma _{k}^{(lm)}(t-\tau )}}
\label{CdeTexpansion}
\end{equation}
and
\begin{align}
& {{C}_{ml}}(\tau ,t)=C_{lm}^{*}(t,\tau ) \nonumber \\ 
& =\sum\nolimits_{k}^{{{K}^{(lm)}}}{\bar{\alpha }_{k}^{(lm)}}{{e}^{i\gamma _{k}^{(lm)}(t-\tau )}}  
 \label{CdeTconjg}
\end{align}
(The possibility of keeping the same exponent in expressions (\ref{CdeTexpansion}) and (\ref{CdeTconjg}) is illustrated in the \modif{supplementary material} \cite{Pomyalov2010}). The time non-local master equation (Eq.(\ref{eq:mastereq})) is replaced by time local coupled equations among auxiliary density operators (ADOs) having the dimension of the reduced system matrix. The initial condition is assumed to be factorized ${{\rho }_{tot}(t=0)}={{\rho }_{S}}(t=0)\rho _{h_0}^{eq}(T)$ where $\rho _{h_0}^{eq}(T)$ is the \modif{equilibrated bath interacting with the ground state} with harmonic bath $h_0$. This is justified for an ultra-short preparation by a Franck-Condon (FC) process that projects the ground equilibrium bath in each excited electronic state \cite{Fleming2010}. The ADOs will describe the behavior of the Gaussian environment. With the FC preparation, they are initially set to zero.  Each auxiliary matrix ${{\tilde{\rho }}_{\mathbf{m}}}(t)$ is denoted by a vector of indices giving the occupation number in each artificial mode of the diagonal and off-diagonal correlation functions:
\begin{equation}
\mathbf{m}=\left( n_{1}^{(11)}..n_{K}^{(11)},..,n_{1}^{(ll)}..n_{K}^{(ll)},..n_{1}^{(lm)}..n_{K}^{(lm)} \right)
\end{equation}
with $l,m\in \left[\!\left[ 1,{{N}_{bath}} \right]\!\right]$ where ${{N}_{bath}}$ is the number of baths in the LVC model. Everything happens as if one considers an extended ensemble of baths. When the uncorrelated case involves ${{N}_{bath}}$  baths with spectral densities ${{J}_{ll}}(\omega )$ and ${{J}_{mm}}(\omega )$, one now adds the baths ${{J}_{lm}}(\omega )$ and ${{J}_{ml}}(\omega )$.  Note that some uncorrelated baths may not contribute to the off-diagonal part, as it is assumed here in the LVC case for the interstate coupling bath. 

\begin{figure}
 \centering
 \includegraphics[width =1\columnwidth]
{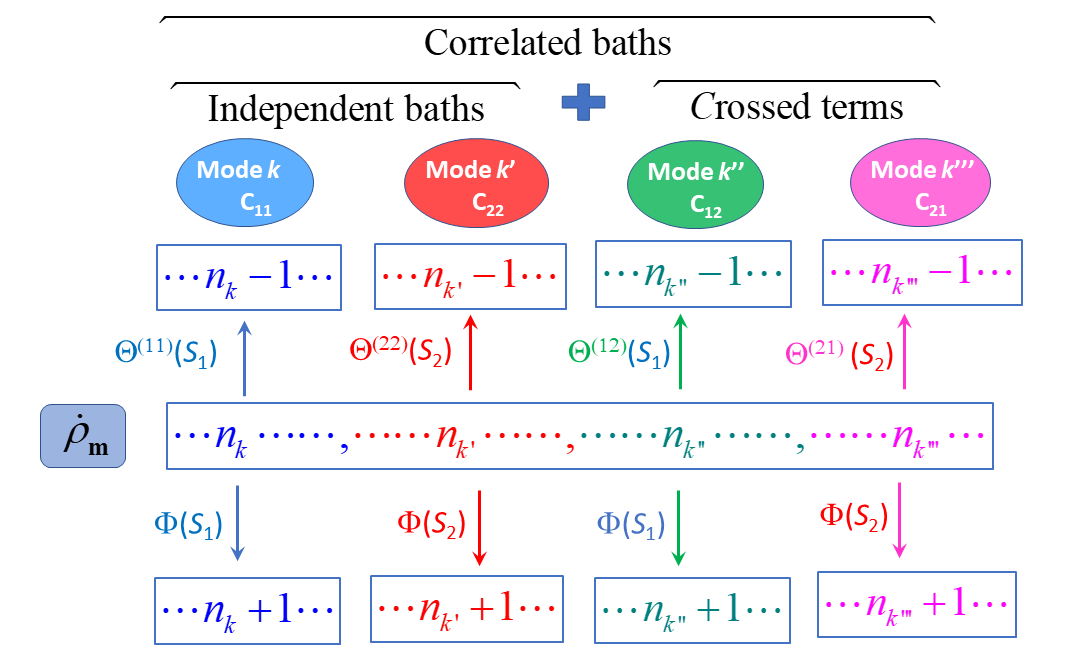}
\caption{Illustration of the ADOs involved in the master equation (Eq.(\ref{eq:HEOMeq})) for the matrix $\dot{\rho}_{\textbf{m}} $ in the two-bath case showing the complexity of the hierarchy for the correlated case with respect to the model with independent baths. $n_k$ is the occupation number in the $k^{th}$ mode of the expansion of the correlation function of each bath or of the crossed correlation function.  ${{\Phi }_{n}}(t)\centerdot =-\left[ {{S}_{n}}(t),\centerdot  \right]$ and $\Theta _{k}^{(lm)}(t)\centerdot =\alpha _{k}^{(lm)}{{S}_{m}}(t)\centerdot -\bar{\alpha }_{k}^{(lm)}\centerdot {{S}_{m}}(t)$. }  
\label{fig:HEOMcorrebaths}
\end{figure}

The coupled equations in the interaction representation are written by using compact operators: 
\begin{align}
  & {{{\dot{\tilde{\rho }}}}_{\mathbf{m}}}(t)=i\sum\limits_{l,m=1}^{{{N}_{bath}}}{\sum\limits_{k=1}^{{{K}^{(lm)}}}{n_{k}^{(lm)}}}\gamma _{k}^{(lm)}{{{\tilde{\rho }}}_{\mathbf{m}}}(t) \nonumber \\ 
 & -\sum\limits_{l=1}^{{{N}_{bath}}}{{{\Phi }_{l}}(t)\sum\limits_{m=1}^{{{N}_{bath}}}{\sum\limits_{k=1}^{{{K}^{(lm)}}}{\tilde{\rho }_{\mathbf{m}_{k}^{+}}^{{}}(t)}}} \nonumber \\ 
 & -\frac{1}{{{\hbar }^{2}}}\sum\limits_{l,m=1}^{{{N}_{bath}}}{\sum\limits_{k=1}^{{{K}^{(lm)}}}{n_{k}^{(lm)}}}\Theta _{k}^{(lm)}\tilde{\rho }_{\mathbf{m}_{k}^{-}}^{{}}(t)
 \label{eq:HEOMeq}
\end{align}
where $\textbf{m}_k^{\pm}={\left\{ ...n^{(lm)}_{k}\pm 1.... \right\}}$ represents the index vector where the occupation number of the $k^{th}$ mode of bath $(lm)$ increases or decreases by one unit while all the other numbers remain unchanged. ${{\Phi }_{l}}(t)\centerdot =-\left[ {{S}_{l}}(t),\centerdot  \right]$ with $l=1,{{N}_{bath}}$ and 	$\Theta _{k}^{(lm)}(t)\centerdot =\alpha _{k}^{(lm)}{{S}_{m}}(t)\centerdot -\bar{\alpha }_{k}^{(lm)}\centerdot {{S}_{m}}(t)$ with $k=1,{{K}^{(lm)}}$. As illustrated in fig.\ref{fig:HEOMcorrebaths}, in the uncorrelated case, the terms ${{\Phi }_{l}}(t)\centerdot $ involves a commutator of the system operator ${{S}_{l}}(t)$ only with raising ADOs of the corresponding bath with correlation function ${{C}_{ll}}(t,\tau )$ while in the cross-correlated case one must add commutators of this ${{S}_{l}}(t)$ with raising terms of the additional baths described by ${{C}_{lm}}(t,\tau )$. Similarly, without cross-correlation, only diagonal terms  $\Theta _{k}^{(ll)}(t)\centerdot $ arise. They involve the parameters of  ${{C}_{ll}}(t,\tau )$ and the system operator ${{S}_{l}}(t)$ that acts on the lowering ADOs of this bath $ll$. With cross-correlation, the supplementary term $\Theta _{k}^{(lm)}(t)\centerdot $ involves the parameters of ${{C}_{lm}}(t,\tau )$ with action on lowering ADOs of the bath $lm$ but with the system operator ${{S}_{m}}(t)$. Similarly, $\Theta _{k}^{(ml)}(t)\centerdot $ implies bath $ml$ and the system operator ${{S}_{l}}(t)$.

It is obvious that taking into account the cross-correlation involves $N_{bath}^2$ so that the number of ADOs rapidly grows and computational resources become prohibitive when the hierarchy level is high. We now revisit and discuss a strategy to reduce the number of baths by introducing approximate uncorrelated shared baths.     

\section{Approximate de-correlation for HEOM } 
\label{sec:Transfo}
\modif{When the baths are not fully or anti-correlated as assumed in sec.\ref{sec:fully/fullyanti}, the central issue in bath de-correlation is whether the transformation used to diagonalize the spectral density matrix is frequency dependent or frequency independent within the HEOM framework. We will show that this frequency dependence could be taken into account in T-TEDOPA. In general, the spectral density matrix $J_{ij}(\omega)$ can always be diagonalized at each frequency, but if the corresponding transformation varies with $\omega$, the bath operators become frequency dependent mixtures of the original operators. In that case, the transformed baths cannot generally be represented by a simple set of independent harmonic environments compatible with the standard HEOM formulation.}

\modif{As a consequence, the practical possibility of bath de-correlation strongly depends on the spectral representation itself and we will distinguish the discrete or continuous representation.} 

\modif{(a) \textit{Discrete case}. For discrete LVC spectral densities on a single site, the correlation matrix is $J_{lm}^{\beta }(\omega_j )$, which is linked to the discrete spectral density $J_{lm}^{{}}(\omega_j )$ (Eq.(\ref{eq:Jbeta})). When the baths are at the same temperature, we may discuss $J_{lm}^{{}}(\omega_j )$ only.
}
 At each  frequency $\omega_j$, the latter is a dyadic matrix 
\begin{equation}
\mathbf{J}({{\omega }_{j}})={{\mathbf{g}}_{j}}\otimes {{\mathbf{g}}_{j}}    
\end{equation}
built from the vectors ${{\mathbf{g}}_{j}^T}=\left( g_{j}^{(1)},..,g_{j}^{({{N}_{bath}})} \right)$ (Eq.(\ref{eq:spectraldensity})). Accordingly, $J_{lm}^{\beta }(\omega_j )$ has a single non-zero eigenvalue
\begin{align}
  & \mathbf{\bar{J}}_{{}}^{\beta }(\omega_j )={{\mathbf{U}^T}}(\omega_j )\mathbf{J}_{{}}^{\beta }(\omega_j )\mathbf{U}(\omega_j ) \nonumber \\ 
 & =diag({{\Lambda }_{1}}(\omega_j ),0..0)  
\end{align}
with ${{\Lambda }_{i}}({{\omega }_{j}})=0,\forall i\in \left[\!\left[ 2,{{N}_{bath}} \right]\!\right]$. In the discrete case, there is only one non-zero eigenvalue $\Lambda_1(\omega_j)$ at each frequency. 

\modif{(b) \textit{Continuous case}. After spectral broadening and possible fit by analytical expressions, however, positive and negative contributions may cancel over finite frequency windows, leading to a much smaller continuous cross spectral density $J_{12}(\omega)$. The resulting diagonalization can therefore differ substantially between the discrete and continuous representations, potentially giving more non-zero eigenvalues or making the baths appear effectively uncorrelated only after smoothing.} A deviation from this ideal dyadic behavior may also occur when the spectral densities are Fourier transforms of correlations functions computed by molecular dynamics \cite{DunnettMB2021,Hunter2024_ie,Zuehlsdorff2024}.

De-correlating the baths is then in principle possible by working with the transformed collective modes and the transformed system operators as done in Sec.\ref{sec:fully/fullyanti}. This would lead to a single effective shared bath associated to the $\Lambda_1$ eigenvalue in the ideal case that preserves the dyadic structure of the discrete matrix. Working with uncorrelated baths and, even better, with a single shared bath, would substantially reduce the computational resources in the HEOM simulations. However, the frequency dependence of the system operators $\bar{\textbf{s}}=\textbf{U}^T(\omega)\textbf{s}$ and of the combination of the bath modes $\bar{\textbf{b}}=\textbf{U}^T(\omega)\textbf{b}$ is not suitable for HEOM. The goal is to find the optimal frequency independent transformation matrix $\textbf{U}_{\text{opt}}$ to minimize the correlations while ensuring a good approximation of the dynamics
\begin{equation}
{\bar{\mathbf{J}}^{\beta}_{{{U}_{\text{opt}}}}(\omega)}=\mathbf{U}^T_{\text{opt}}\mathbf{J^{\beta}(\omega )}{{\mathbf{U}}_{\text{opt}}}. 
\label{eq:Jtransfoopt}
\end{equation}
The quality of $\textbf{U}_{\text{opt}}$ may be estimated by the remainder of the off-diagonal terms
\begin{equation}
\varepsilon ({{\mathbf{U}}_{\text{opt}}})=\int{d\omega }{{\sum\limits_{i\ne j}{\left| {{\left( \mathbf{U}_{\text{opt}}^T{{\mathbf{J}}^{\beta }}(\omega ){{\mathbf{U}}_{\text{opt}}} \right)}_{ij}} \right|}}^{2}}
\label{eq:error}
\end{equation}
Different possibilities may be examined:

(a) The approximate constant $\textbf{U}_{\text{opt}}$ matrix  is denoted with an index $\textbf{U}_{\eta}$ when it is chosen at a representative frequency, for instance that of the dominant peak in the spectral density. This assumes that the spectral densities keep the same shape with a constant ratio among the couplings. For instance, in the two-bath model ${{g}^{(2)}}(\omega )=\pm \sqrt{\eta }{{g}^{(1)}}(\omega ) \space \,\forall \omega  $ with $\eta > 0$ and a constant sign. The cases where $\sqrt{\eta }= 1$ are the fully correlated (if ${{g}^{(2)}}(\omega )= + {{g}^{(1)}}(\omega )$) or fully anti-correlated (if ${{g}^{(2)}}(\omega )= - {{g}^{(1)}}(\omega )$) baths, as discussed in Sec.\ref{sec:fully/fullyanti}.  

(b) Another choice, referred to as $\textbf{U}_\text{aver}$ does not postulate the constraint of a constant ratio between the couplings. The matrix is built from the average value of the eigenvectors of $\mathbf{{J}}_{{}}^{\beta }(\omega )$. 

(c) A third possibility is to take $\textbf{U}_{C(0)}$  given by the eigenvectors of the initial value of the correlation matrix $\mathbf{C}(t=0)$. Indeed, each element is an average of the corresponding  matrix element of $\mathbf{J}^{\beta }(\omega )$ since it is its integral as a result of the Fourier transform relation (Eq.(\ref{eq:TFC(t)new})) .    

(d) Finally $\textbf{U}_{\text{opt}}$ may be estimated by $\textbf{U}_\text{PCA}$ obtained by the principal component analysis (PCA) by making the singular value decomposition (SVD) of the $N_{bath} \times N_{bath} \times N_{\omega}$ tensor where $N_{\omega}$ is a large sampling of the frequency domain. 

\begin{figure}
 \centering
\includegraphics[width =0.8\columnwidth]{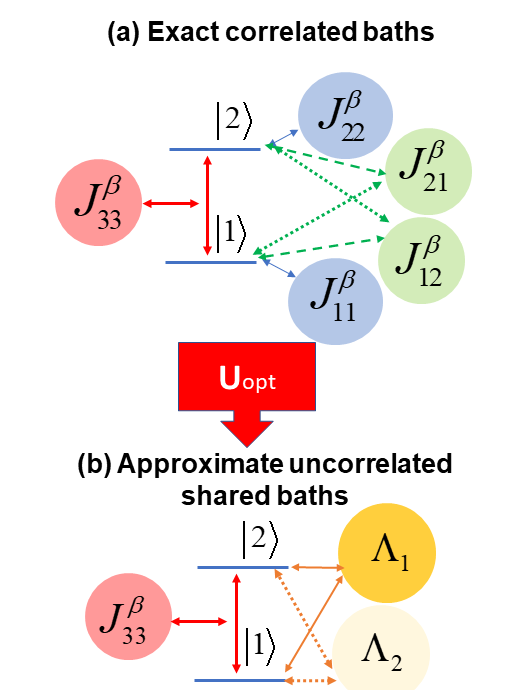}
\caption{Spectral densities of a two-level model with two correlated tuning baths ($J_{11}^{\beta}$, $J_{22}^{\beta}$) diagonally coupled and a coupling uncorrelated bath ($J_{33}^{\beta})$ off-diagonally coupled. Panel (a): Correlated baths. The dashed arrows represent the action of $\Phi$ and the dotted lines that of $\Theta$ in HEOM equations (Eq.(\ref{eq:HEOMeq})). Panel (b): Shared baths after the transformation of the two correlated baths by an approximate frequency independent transformation $\mathbf{U}_{\text{opt}}$ (Eq.(\ref{eq:Jtransfoopt})). In the ideal case, $\Lambda_1(\omega) \gg \Lambda_2(\omega)$ so that $\Lambda_2(\omega)$ may be discarded.}  
\label{fig:modelcorre}
\end{figure}

The exact and approximate strategies for a HEOM simulation are schematized in fig.\ref{fig:modelcorre} for a model discussed below where two correlated tuning baths are diagonally coupled and one uncorrelated coupling bath is off-diagonally coupled to the excited states. The exact simulation involves the HEOM equations with correlated baths (Eqs.(\ref{eq:HEOMeq})), illustrated in fig.\ref{fig:modelcorre}~(a). In some favorable cases, it can be followed with an approximate diagonalization of $\mathbf{J}^{\beta }(\omega )$ by a frequency independent $\mathbf{U}_{\text{opt}}$ matrix that provides two decoupled shared baths described by \modif{the new spectral functions} $\Lambda_1(\omega)$ and $\Lambda_2(\omega)$ with in the ideal case $\Lambda_1(\omega) >> \Lambda_2(\omega)$ so that $\Lambda_2(\omega)$ may be discarded (fig.\ref{fig:modelcorre}~(b)).

\modif{In Eq.(\ref{eq:error}), we have estimated the validity of the approximate frequency independent transformation by the integrated Frobenius projection error. For a deeper estimation of the error, it could be interesting to generalize the error bound proposed in ref.\cite{Mascherpa2017} to quantify the impact on the expectation value of observables due to a variation of a single spectral density $\Delta J(\omega)$. The error is expected to grow exponentially with time and is quantified by parameters that are related the correlation kernel (Eq.(\ref{eq:TFC(t)new})) by transforming $\Delta J(\omega)$ in place of $J(\omega)$. In the case of multiple correlated baths, the residual correlation matrix $\mathbf{J^{\beta}(\omega )-{\bar{\mathbf{J}}^{\beta}_{{{U}_{\text{opt}}}}(\omega)}}$ (Eq.(\ref{eq:Jtransfoopt})) is the source of error and the integrated strength of this residual bath correlation kernel derived from Eq.(\ref{eq:TFC(t)new}) could be involved in the error estimation as in the single-bath result.}   

\section{T-TEDOPA for correlated baths}
\label{sec:T-TEDOPA}
By the efficient T-TEDOPA mapping of Tamascelli \textit{et al}.\cite{Tamascelli2019,Tamascelli2020}, the initial thermalized mixed state is replaced by a pure state of the environment at 0 K. The initial factorized state becomes:
\begin{equation}
\rho \left( t=0 \right)\text{ }={{\rho }_{S}}\left( t=0 \right)\otimes  \ket{0 \ldots 0}  \bra{0 \ldots 0}
\end{equation}
where $\ket{0 \ldots 0}$  is the state with all the environmental modes in their ground state. The modes are defined by sampling the temperature-dependent spectral density $J^{\beta}(\omega)$ on the extended frequency domain comprising negative frequencies. In the basic case where a given electronic state $n$ is coupled to a unique bath by the coupling operator ${{S}_{n}}=\left| n \right\rangle \left\langle  n \right|$, the system bath interaction in the continuous frequency domain may be written:
\begin{equation}
{{H}_{SB}}=\int{d\omega\sqrt{{{J}^{\beta}_{nn}}(\omega )}}{{S}_{n}}q_{B_n}(\omega ).    
\end{equation}
The infinite bath is mapped onto a discrete set of chain
modes by an unitary transformation built with polynomials orthogonal with respect to the measure of the spectral density $J_{nn}^{\beta}(\omega)d\omega$ \cite{Chin2010}. The system interacts only with the first mode of the chain. All the coupling coefficients connecting the neighboring modes are obtained  by a procedure based on the three terms recurrence relation of the polynomials \cite{Chin2010,deVega2015,Dunnett2021}. The length of the chain must be adapted in order to avoid the return of the excitation to the system. The possible excitation in each mode is fixed by the Fock space of dimension $d$. This configuration leads to the MPS representation formed by a chain of tensor $d$-dimensional cores bounded by the bond dimension $r$. 

\begin{figure}
 \centering
\includegraphics[width =1.\columnwidth]{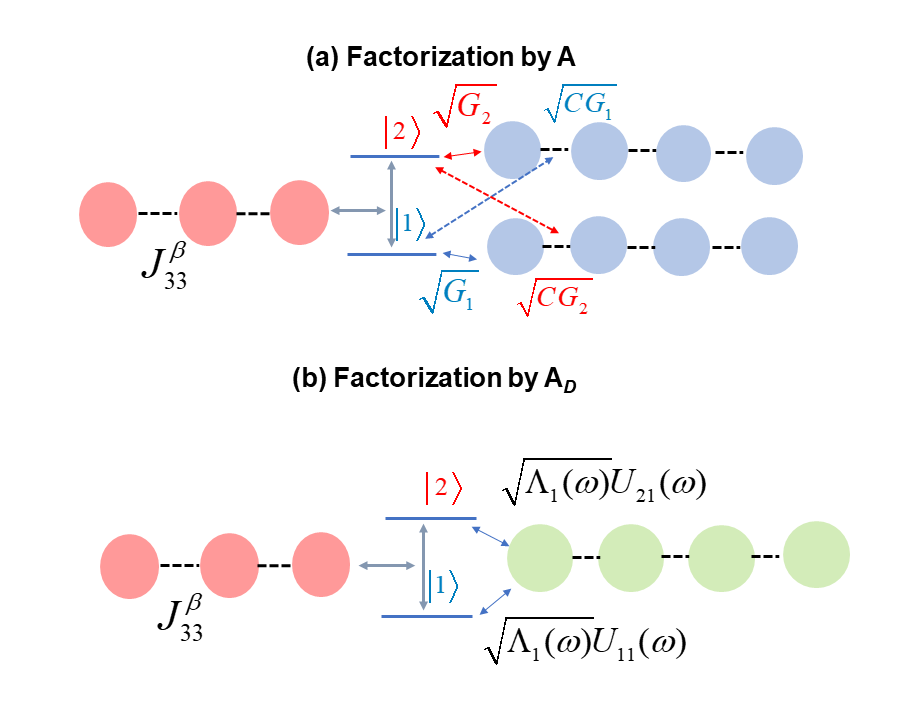}
\caption{Scheme of the coupling linear chains in the T-TEDOPA method for the different factorizations of the $\textbf{J}^{\beta}(\omega)$ matrix. (a) Transformation of $H_{SB}$ with the $\textbf{A}(\omega)$ matrix (Eq.(\ref{eq:matA})). Each state is coupled to the two baths ensuring a significant partial decorrelation. (b) After the frequency dependent transformation with the $\textbf{A}_D(\omega)$ matrix (Eq.(\ref{eq:matAD})) providing a complete de-correlation, a single shared chain subsists in the ideal case when the dyadic structure of the density matrix is preserved. } 
\label{fig:modelMPS}
\end{figure}

When the system interacts with two correlated tuning baths, a transformation of the $H_{SB}$ operator has been proposed recently \cite{DunnettMB2021,Hunter2024_ie,Zuehlsdorff2024}. This transformation uses a particular matrix $\textbf{A}(\omega)$  that factorizes the spectral density as followed:
\begin{equation}
{{\mathbf{J}}^{\beta }}(\omega )=\mathbf{A}(\omega ){{\mathbf{A}}^{T}}(\omega ).    
\end{equation}
The $\textbf{A}(\omega)$ matrix transforms the coupling operators $\bar{\textbf{s}}_A(\omega)=\textbf{A}(\omega)\textbf{s}$ and the bath modes $\bar{\textbf{b}}_A=(\textbf{A}^{T})^{-1}\textbf{b}$. The choice of the factorization matrix $\textbf{A}$ is not unique. In this proposal, the $\textbf{A}$ matrix is (by dropping the writing of $\omega$):
\begin{equation}
\mathbf{A}=\left( \begin{matrix}
   \sqrt{{{G}_{1}}} & \sqrt{C{{G}_{2}}}  \\
   \sqrt{C{{G}_{1}}} & \sqrt{{{G}_{2}}}  \\
    \end{matrix} \right)
\label{eq:matA}
\end{equation}
with
\begin{equation}
{{G}_{1}}={{\left( J_{12}^{\beta } \right)}^{2}}/\left[ 2J_{22}^{\beta }(1-R) \right],   
\end{equation}
\begin{equation}
 {{G}_{2}}={{\left( J_{12}^{\beta } \right)}^{2}}/\left[ 2J_{11}^{\beta }(1-R) \right],   
\end{equation}
\begin{equation}
C=(2J_{11}^{\beta }J_{22}^{\beta }/{{(J_{12}^{\beta })}^{2}})(1- R)-1,    
\end{equation}
 and $R={{\left( 1-{{(J_{12}^{\beta })}^{2}}/(J_{11}^{\beta }J_{22}^{\beta }) \right)}^{1/2}}$. The two baths are independent and each bath couples to the two states by different spectral densities as illustrated in fig.\ref{fig:modelMPS}
 \begin{align}
  & {{H}_{SB}}=\int{d\omega }\left( {{A}_{11}}(\omega ){{S}_{1}}+{{A}_{12}}(\omega ){{S}_{2}} \right){{q}_{{{B}_{A1}}}}(\omega ) \nonumber \\ 
 & +\int{d\omega }\left( {{A}_{21}}(\omega ){{S}_{1}}+{{A}_{22}}(\omega ){{S}_{2}} \right){{q}_{{{B}_{A2}}}}(\omega ).  
\end{align}
For each bath, the chain mapping is done with respect to $G_1(\omega)$ and $G_2(\omega)$ respectively. As a result, interactions between the other state and the chain modes may become long-ranged \cite{DunnettMB2021,Hunter2024_ie}.

 Inspired by the procedure relevant in HEOM, one could propose to use the factorization with the orthogonal matrix built with the spectral decomposition of $\mathbf{J}(\omega)$ and the diagonal matrix $\mathbf{\Lambda }(\omega )$
 \begin{align}
  & \mathbf{J}_{{}}^{\beta }(\omega )=\mathbf{U}(\omega )\mathbf{\Lambda }(\omega ){{\mathbf{U}^T}}(\omega ) \nonumber \\ 
 & ={{\mathbf{A}}_{D}}(\omega )\mathbf{A}_{D}^{T}(\omega)  
\end{align}
with
 \begin{equation}
 \textbf{A}_D(\omega)=\textbf{U}(\omega)\sqrt{{\mathbf{\Lambda}}(\omega)}. 
 \label{eq:matAD}
 \end{equation}
 
By using $ \textbf{A}_D(\omega)$ for the transformation, the system operator becomes $\bar{\textbf{s}}_{A_D}(\omega)=\textbf{A}_D(\omega)\textbf{s}$. One finds a similar scheme with two uncorrelated baths, each of which is coupled to the two states with different spectral densities. However, in a favorable case when the dyadic structure of the spectral density matrix is nearly conserved, a single eigenvalue $\Lambda_1(\omega)$ dominates ($\Lambda_1(\omega)>>\Lambda_2(\omega)$) and a single shared bath might be sufficient with 
\begin{equation}
{{H}_{SB}}=\int{d\omega }\left( {{c}_{1}}(\omega ){{S}_{1}}+{{c}_{2}}(\omega ){{S}_{2}} \right){{q}_{{{B}_{A_{D1}}}}}(\omega ) 
\label{eq:HSBTEDOPA}
\end{equation}
where ${{c}_{n}}(\omega )=\sqrt{\Lambda_{1}(\omega )}U_{n1} (\omega )$ ($n=1,2$) and ${q}_{{{B}_{A_{D1}}}}$ represents the new bath coordinate. The eigenvalue is included in the system operator while it is in the transformed correlation function in HEOM.  It is worth noting that the T-TEDOPA procedure allows to account for the full frequency dependence of the transformation $\textbf{U}(\omega)$ at the cost of the long-range interactions. It is an interesting alternative to exact correlated HEOM simulations mainly when a single eigenvalue of the spectral density matrix dominates. If $\Lambda_2(\omega)$ is not negligible, a second term must be added in Eq.(\ref{eq:HSBTEDOPA}) and the procedure becomes similar to the other ones using $\textbf{A}(\omega)$ with two baths interacting with the two states.  However, it is not obvious which of the transformations $\textbf{A}$ or $\textbf{A}_D$ is the best. In principle, the spectral transformation with $\textbf{A}_D$ de-correlates the baths. The $\textbf{A}$ procedure also uses independent baths but with a rotation of the spectral modes. The $\textbf{A}$ matrix may be written $\textbf{A}=\mathbf{U}\sqrt{\mathbf{\Lambda}}\mathbf{O}$ where  $\mathbf{O}$ is an orthogonal matrix. This combination of spectral modes may induce some dynamical correlation. The length of the chains depends on the shape of the coefficients of the $S_n$ operators provided by the different transformations. It seems that slowly varying functions are more likely to avoid long chains. A kind of compromise must be
found between the complete de-correlation of the
baths and the efficiency of the chains.

We will illustrate that, at least in favorable cases, the factorization with $\textbf{A}_D(\omega)$ giving a single shared bath is possible. In HEOM, the only way to avoid the full correlated simulation is the choice of an optimal frequency independent transformation $\textbf{U}_{opt}$. T-TEDOPA could in principle treat the full correlated case by using the factorization with $\textbf{U}(\omega)\sqrt{\mathbf{\Lambda}(\omega)}$. A simplification may also be used by choosing an optimal frequency independent $\textbf{U}_{opt}$ matrix. This approximation has already been shown in ref.\cite{LeDe2024}. 

\section{ Correlation in a LVC model of a conical intersection}
\label{sec:LVCmodel}
We consider a molecular model with a ground and two excited states coupled by a conical intersection, the ground state being decoupled from the excited manifold. In a diabatic representation, the generic LVC model Hamiltonian centered at the Franck-Condon Point (FCP) geometry reads :

\begin{align}
  & H=\left( \begin{matrix}
   {{\varepsilon}_{0}} & 0 & 0  \\
   0 & {{\varepsilon}_{1}} & 0  \\
   0 & 0 & {{\varepsilon}_{2}}  \\
\end{matrix} \right)+\left( \begin{matrix}
   {{h}_{0}} & 0 & 0  \\
   0 & {{h}_{1}} & 0  \\
   0 & 0 & {{h}_{2}}  \\
\end{matrix} \right) \nonumber \\ 
 & +\left( \begin{matrix}
   0 & 0 & 0  \\
   0 & \sum\nolimits_{j}{\kappa _{j}^{(1)}{{q}_{j}}} & \sum\nolimits_{j}{{{\kappa}'_{j}}{{q}_{j}}}  \\
   0 & \sum\nolimits_{j}{{{\kappa}'_{j}}{{q}_{j}}} & \sum\nolimits_{j}{\kappa _{j}^{(2)}{{q}_{j}}}  \\
\end{matrix} \right)  
\label{eq:LVCmodel}
\end{align}
where ${{\varepsilon}_{n}}$ ($n=0,1,2$) are the diabatic potential energies at the FCP for each normal mode.  The vibrational Hamiltonian of electronic state is supposed to be harmonic. By neglecting any Duschinsky rotation, the normal modes are assumed to be the same in each electronic state.  Only their equilibrium position $q_{j,eq}^{(n)}$ is different.  In this harmonic approximation, the excited electronic energies at FCP ${{\varepsilon}_{n}}={{\varepsilon }_{n}^0}+{{\lambda }_{n}}$ ($n=1,2$) are the sum of the energy at the minimum for each mode $q_{j,eq}^{(n)}$ and the renormalization energy  ${{\lambda }_{n}}=\frac{1}{2}\sum\nolimits_{j}{\omega _{j}^{2}}q_{j,eq}^{(n)2}$. $\kappa _{j}^{(n)}=\omega _{j}^{2}\left( q_{j,eq}^{(0)}-q_{j,eq}^{(n)} \right)$ are the gradients of the diabatic potential energy curves at the FCP and the sum runs over all the tuning modes. ${{\kappa}'_{j}}$ are the gradients of the electronic interstate coupling.  The diagonal and off-diagonal sums define collective modes having the dimension of an energy: the collective tuning modes $B_{n}^{{}}=\sum\nolimits_{j}^{{}}{\kappa _{j}^{(n)}q_{j}^{{}}}$ ($n=1,2$) that are coupled to the system by $S_n=\ket{n} \bra{n}$ and the interstate coupling collective mode ${{B}_{3}}=\sum\nolimits_{j}^{{}}{{{\kappa}'_{j}}}q_{j}^{{}}$ coupled by $S_3=\ket{1}\bra{2} + \ket{2}\bra{1}$. The two baths $B_1$ and $B_2$ are correlated as illustrated in Sec.\ref{sec:fully/fullyanti} for the fully correlated case. We assume that the coupling bath $B_3$ is completely uncorrelated. It is here a simple spectator bath for the correlation problem but it plays a significant role in the non-adiabatic dynamics.

\section{Illustrative Application}
\begin{figure}
 \centering
\includegraphics[width =1.\columnwidth]{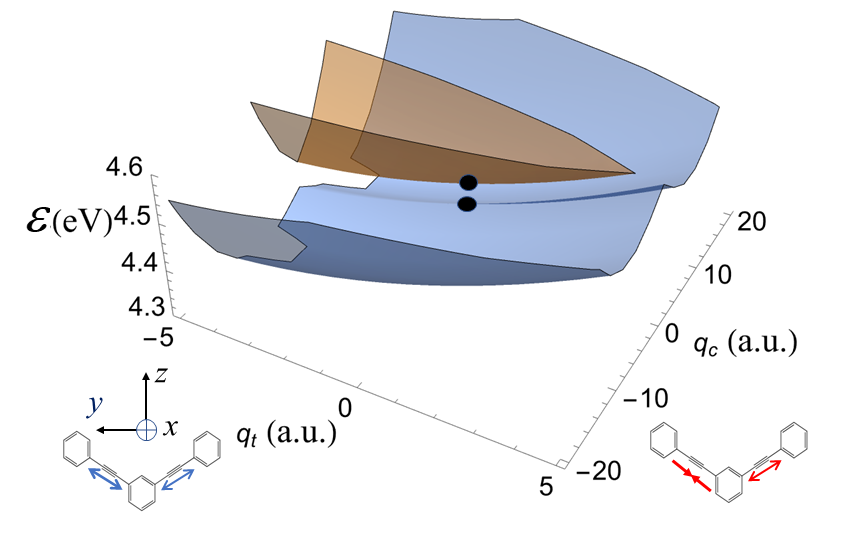}
\caption{Illustration of a situation suitable for positive correlated baths. 2D adiabatic potential energy surfaces of the $\text{S}_1$ and $\text{S}_2$ states of the 3-bis(phenylethynyl)benzene (referred to as m22) along the main high frequency tuning and coupling modes that are the symmetrical and antisymmetric stretching vibrations of the two acetylenic bonds. The mass-weighted coordinates $q_t$ and $q_c$ are given in atomic units ($\sqrt{m_e}a_0$). \modif{The black circles indicate} the FC positions where the gradients have the same sign.   }  
\label{fig:modelLVC}
\end{figure}

We present an illustrative example based on the $\textit{ab}$ $\textit{initio}$ LVC model of the 1,3-bis (phenylethynyl)benzene dimer where a conical intersection couples the excited states $\text{S}_1$ and $\text{S}_2$ (see fig.\ref{fig:modelLVC}). \modif{The corresponding potential energy surfaces and inter-state coupling have been computed using DFT (ground state) and TD-DFT (excited states) at the CAM-B3LYP/6-31$+$G$*$ level of theory calculated by B. Lasorne and coworkers \cite{Ho2019,Galiana2023}. } \modif{The minimum-energy conical intersection (MECI) of the seam was characterized with the branching-space vectors associated to the halved gradient difference vector (GD) and the derivative coupling vector (DC).} This is referred to as the m22 dimer in the following. \modif{All LVC data (gradients along the normal modes at the FC geometry) have already been used in previous simulations \cite{Galiana2023,Jaouadi2022} and are given in the supplementary material of these works.}

This is an example where the tuning baths obtained from Eq.(\ref{eq:LVCmodel}) are strongly correlated and for which the approximate treatment of a shared bath has been used in previous HEOM applications \cite{Jaouadi2022,LeDe2024} without really justifying it. The LVC Hamiltonian is built by retaining only the planar geometry. At the FCP, the two $\text{S}_1$ and $\text{S}_2$ states are very close, at 4.43 and 4.47 eV. The tuning and coupling modes have the A$_1$ or B$_2$ symmetry of the C$_{2v}$ point group, respectively. The two states may be excited via orthogonal transition dipoles, $\mu_y = 3.96$ $ea_0$ for $\text{S}_1$ and $\mu_z = -1.87$ $ea_0$ for $\text{S}_2$ (the axes are represented in fig.\ref{fig:modelcorre}).  

\begin{figure}
 \centering
\includegraphics[width =1.\columnwidth]{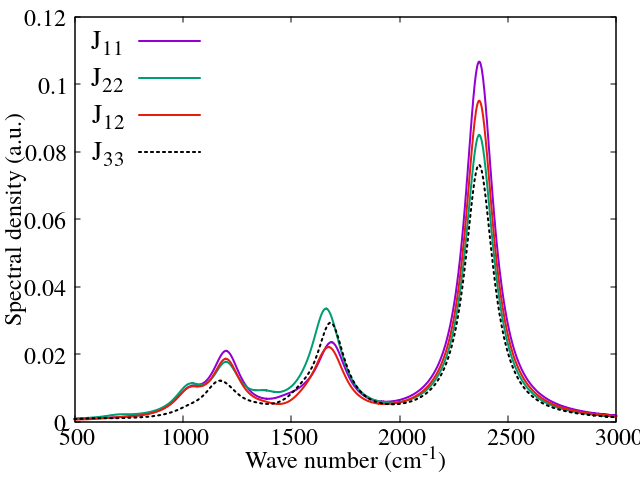}
\caption{Continuous spectral densities in a.u.(Hartree) calibrated from the LVC data of S$_1$ and S$_2$ excited states of the m22 dimer \cite{Galiana2023}. $J_{11}$ and $J_{22}$ are the spectral densities of the tuning baths formed by the symmetrical vibrations. $J_{12}$ is the corresponding crossed term. $J_{33}$ is the spectral density of the interstate coupling bath generated by the antisymmetric vibrations.} 
\label{fig:spectraldensity}
\end{figure}

The continuous spectral densities of the tuning ($J_{11}$, $J_{22}$) and coupling ($J_{33}$) baths and the crossed spectral density ($J_{12}$) are given in fig.\ref{fig:spectraldensity}. In the absence of information provided by molecular dynamics in a solvent, we use a smoothing strategy \cite{Burghardt2010,Burghardt2019} by broadening each discrete data by a Lorentzian function
\begin{equation}
\delta (\omega -{{\omega }_{j}})\sim\frac{1}{\pi }\frac{\Gamma}{{{\left( \omega -{{\omega }_{j}} \right)}^{2}}+{{\Gamma }^{2}}}.
\label{eq:delta}
\end{equation}
The phenomenological width $\Gamma={80}$ cm$^{-1}$ is chosen to match the spectral enlargement introduced in a previous MCTDH simulation of linear spectra \cite{Galiana2023}. For the HEOM simulations, the spectral densities are fitted here by Tannor-Meier Lorentzian functions\cite{Meier1999,Mangaud2023,LeDe2024,Tokieda2025}  that provide an analytical expression of the parameters in the $C_{lm}(t)$ expansion (Eqs.(\ref{CdeTexpansion}) and (\ref{CdeTconjg})). The main relations are summarized in the Supplemental Material.  The dominant peaks arise at the high frequency symmetrical or antisymmetric stretching modes of the acetylenic bond. At this frequency, the gradients have the same sign, which explains the large positive value of the crossed term. After smoothing, the crossed spectral density $J_{12}(\omega)$ has no negative value in this case. Note that the convolution procedure effectively provides negative values when most gradients have different signs. The positively correlated tuning baths are related to what is generally referred to as a sloped conical intersection. 
  
We will compare the two exact simulations at T $=$ 298 K driven by the correlated HEOM equations (Eq.(\ref{eq:HEOMeq})) and by the de-correlated T-TEDOPA method using the factorization by the $\textbf{A}_D(\omega)$ matrix (Eq.(\ref{eq:matAD})). For HEOM, \modif{the main constraints is the very fast increase of the number of ADOs with the number of decay modes of all the correlation functions. It is given by ${{N}_{\text{ADO}}}=\frac{\left( L+K \right)!}{L!K!}$. Five correlation functions are involved in this application, four for the two correlated tuning bath and one for the tuning bath, $K=\sum\nolimits_{l,m=1}^{2}{{{K}^{(lm)}}}+{{K}^{(33)}}$} (see fig.\ref{fig:modelcorre}). Each spectral density is fitted by three Lorentzian functions, each of which \modif{leads} to two decay modes in the correlation function. The contribution of the Matsubara terms coming from the Bose function may be fitted by two exponential functions for each correlation function, as done in refs. \cite{LeDe2024,Lambert2019} (the parameters of this fit at 298 K are given in the SI). These terms are correctly taken into account in the approximate treatment using $\mathbf{U}_{opt}$ but not in the correlated case (Eqs.(\ref{eq:HEOMeq})). \modif{In our simulation of the correlated case, the number of artificial modes is thus $K^{(lm)}=3 \times 2= 6$ (number of Lorentzian functions per bath times number of decay modes per Lorentzian).} At level $L = 7$ of the hierarchy, the number of ADOs generated by the total number of artificial modes \modif{$K=5 \times 6 =30 $} (number of baths times number of modes per bath) is already \modif{$N_{\text{ADO}} = 10,295,472$}. Note that the approximate method with a single tuning shared bath requires significantly less computational effort. \modif{The number of correlation functions is then only 2, one for the shared bath and one for the tuning bath. With 6 decay modes per bath, a simulation at the same level $L = 7$ demands only $N_{\text{ADO}} = 50,388$ } \modif{This reduction of the computational resource } allows us to take into account the Matsubara terms by the fitting procedure. \modif{With two supplementary functions per bath, the number of modes is now 8 and at level $L = 7$}, which involves $N_{\text{ADO}} = 116,280$. \modif{Note that a filtering procedure could reduce the ADO numbers during the dynamics \cite{Shi2014,Jaouadi2022}.} \modif{ T-TEDOPA} dynamics is done with a rank $r=8$, a Fock space $d=10$, 100 modes for the coupling bath and 90 modes for the tuning bath(s). \modif{Dynamics with the $A_D$ procedure involves a single shared tuning bath. The storage then requires $N_{TEDOPA}= (100+90) \times 10 \times 8^2= 121,600$ (total number of modes times the dimension of the Fock space times the square of the rank) complex elements. In the $A$ approach, there are two tuning baths and this involves $N_{TEDOPA}= (100+180) \times 10 \times 8^2= 179,200$ elements. In terms of storage resources, the T-TEDOPA method is clearly more effective than HEOM (discussed here in the standard implementation that could be notably improved by the tensor train method). A comparison among different implementations was already made in ref. \cite{LeDe2024}. The $A_D$ strategy is also more advantageous, at least when the density matrix has only one non-zero eigenvalue. Even if the comparison of computational time strongly depends on the hardware used, in our simulations, T-TEDOPA was always more efficient in computational time.  } 
   
\begin{figure}
 \centering
\includegraphics[width =1.\columnwidth]{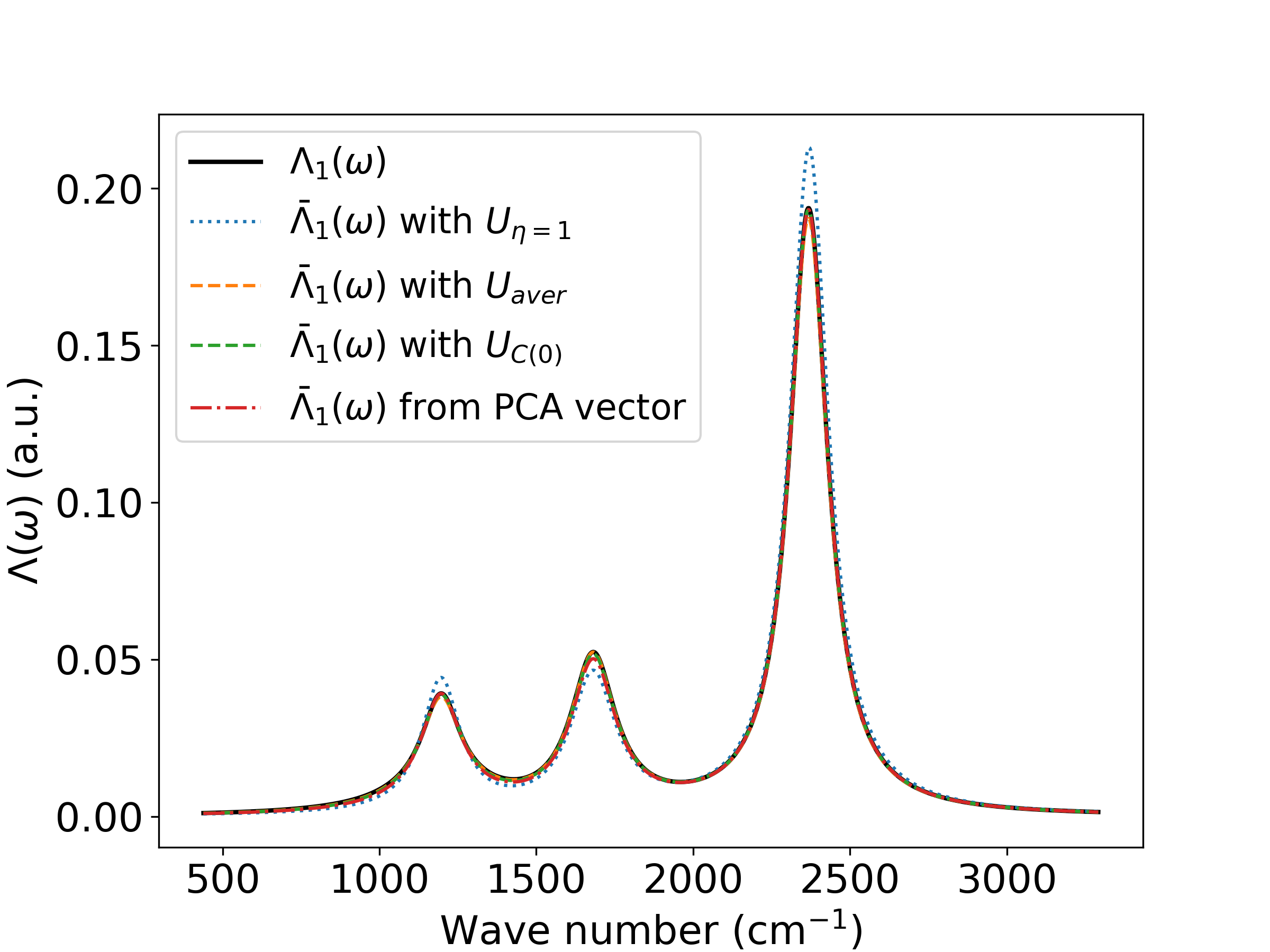}
\includegraphics[width =1.\columnwidth,height=6cm]{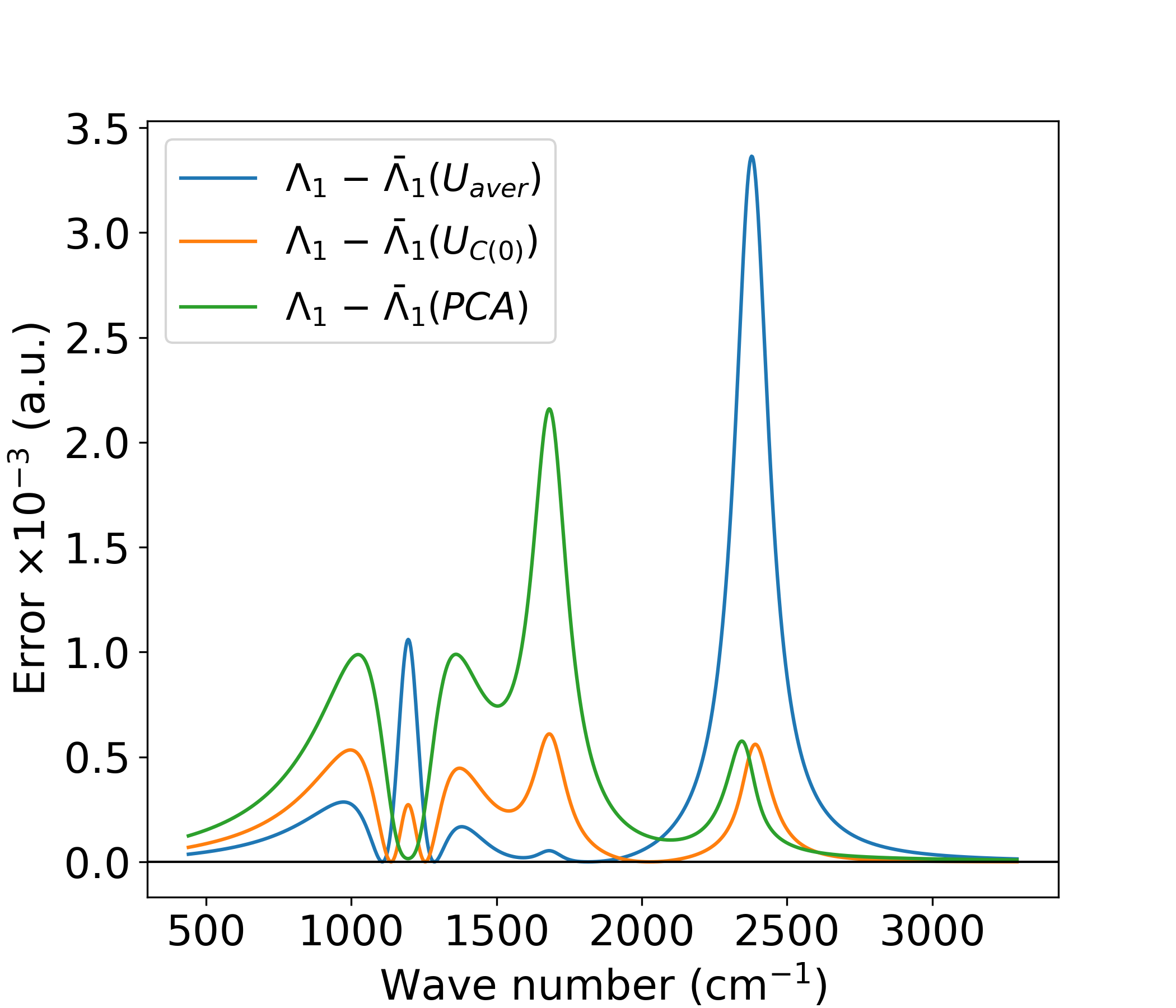}
\caption{Upper panel: Eigenvalue $\Lambda_1(\omega)$ of the continuous $\textbf{J}^{\beta}(\omega)$ of the m22 dimer at 298 K after the Lorentzian smoothing (Eq.(\ref{eq:delta})) and the approximate non-zero eigenvalue $\bar{\Lambda}_1(\omega)$ obtained with a frequency independent $\textbf{U}_{opt}$ matrix described in sec.\ref{sec:Transfo}.  Lower panel: the corresponding errors }   
\label{fig:VPcorre}
\end{figure}

{The demanding resources for solving the correlated HEOM at high level stimulate the use of the approximate shared bath when this is possible. Obviously, the ratio between the two spectral densities $J_{22}(\omega)/J_{11}(\omega)$ depends on the frequency and one has to choose the best frequency independent transformation $\textbf{U}_{opt}$ for the approximate de-correlation.  The eigenvector components $U_{n1}(\omega)$ ($n = 1,2$) of $\textbf{J}^{\beta}(\omega)$ associated with the dominant eigenvalue $\Lambda_1(\omega)$ are shown in the upper part of fig.\ref{fig:VecPcorre} ($\Lambda_2(\omega)$ is negligible). The lower panel of fig.\ref{fig:VecPcorre} shows the elements of the $\textbf{A}_D(\omega)$ (Eq.(\ref{eq:matAD})) and $\textbf{A}(\omega)$ (Eq.(\ref{eq:matA})) matrices used in the T-TEDOPA applications. 

\begin{figure}
 \centering
\includegraphics[width =1.\columnwidth]{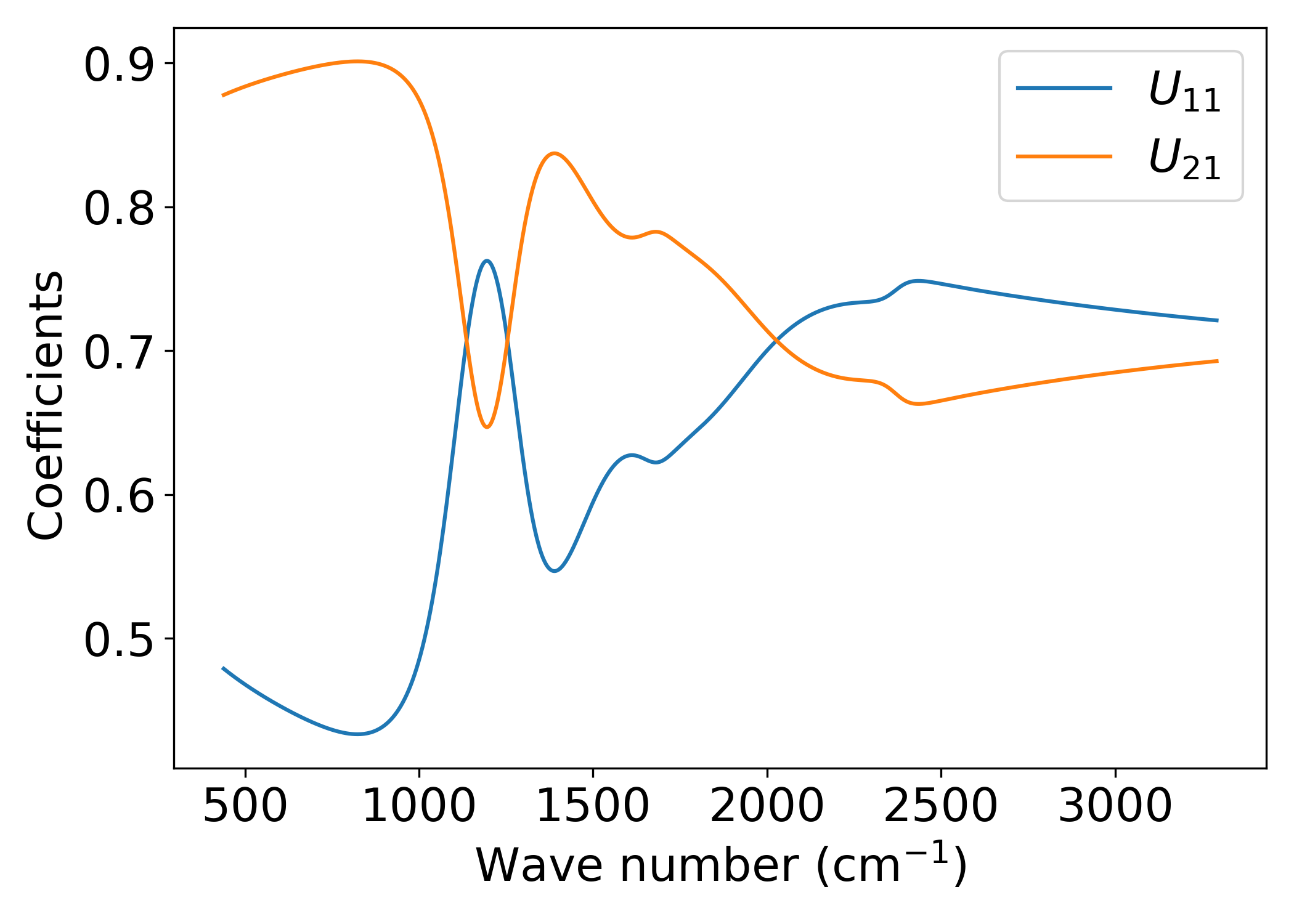}
\includegraphics[width =1.\columnwidth]{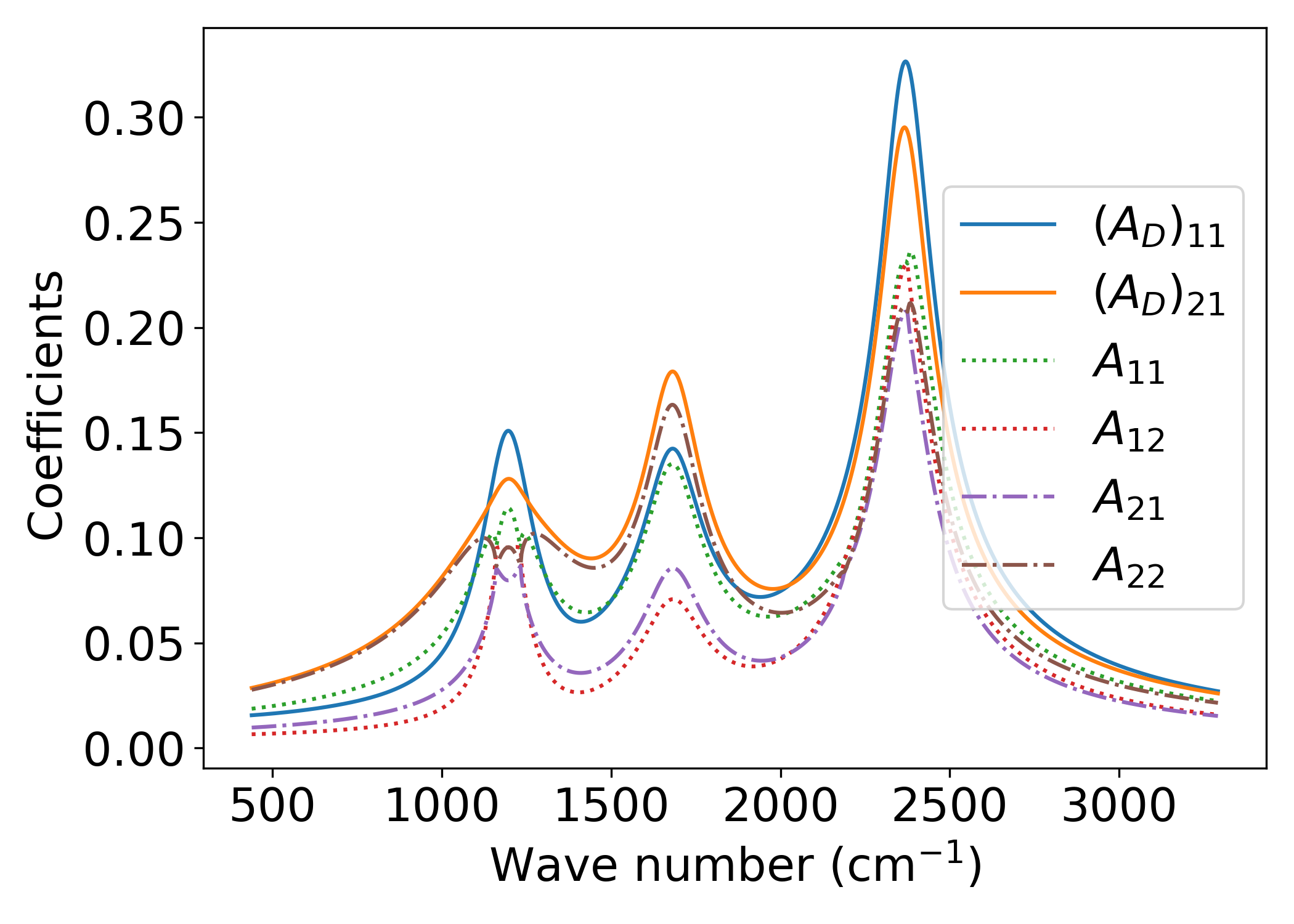}
\caption{Upper panel: coefficients $U_{n1}(\omega)$ ($n = 1,2$) of the transformed operator $\bar{S}_1(\omega)=U_{11}(\omega)S_1+U_{21}(\omega)S_2$ associated with the eigenvalue $\Lambda_1(\omega)$ of $\textbf{J}^{\beta}(\omega)$ for the m22 dimer at 298 K ($\Lambda_2(\omega)$ is negligible). Lower panel: Solid lines: coefficients of the $S_1$ and $S_2$ operators ($\sqrt{\Lambda_1(\omega)}U_{n1}(\omega)$) using the spectral transformation $\textbf{A}_D(\omega)$ (Eq.\ref{eq:matAD}) in TEDOPA. Dots and dash-dots: coefficients of the operators when the transformation is $\textbf{A}(\omega)$ (Eq.\ref{eq:matA}).}
\label{fig:VecPcorre}
\end{figure}

Figure \ref{fig:VPcorre} gives the dominant eigenvalue $\Lambda_1(\omega)$ of $\textbf{J}^{\beta}(\omega)$. The second eigenvalue $\Lambda_2(\omega)$ is negligible. We compare this dominant eigenvalue with the approximate values $\bar{\Lambda}_1(\omega)$ obtained by the different strategies searching for a constant transformation matrix $\textbf{U}_{opt}$ presented in sec.\ref{sec:Transfo}. In the strategy based on the the projection of the principal vector of the PCA, we use a tensor of 1000 vectors $(J^{\beta}_{11}(\omega_k),J^{\beta}_{22}(\omega_k),2J^{\beta}_{12}(\omega_k))$ with a regular sampling and the main PCA vector maximizes the variance. The errors (Eq.(\ref{eq:error})),} which should vanish with the exact $\textbf{U}(\omega)$ matrix, are very small $5.07 \times 10^{-5}$ for $\textbf{U}_\text{aver}$ (strategy (b)), $3.28 \times 10^{-5}$ for $\textbf{U}_{C(0)}$ (strategy (c)) and $4.51 \times 10^{-5}$ for $\textbf{U}_\text{PCA}$ (strategy (d)).

In $\textbf{U}_{\eta =1}$ the coefficients $U_{11}$ and $U_{21}$ are equal to 0.707106 while in $\textbf{U}_{C(0)}$, $U_{11}=0.70707 $ and $U_{21} = 0.70714$, which are roughly the same. This approximation with $\textbf{U}_{\eta =1}$ is an acceptable approximation in this example. It corresponds to the well known transformation of the modes allowing to describe the conical intersection in terms of three effective modes. The first with linear couplings $\kappa _{j}^{+}=(\kappa _{j}^{(1)}+\kappa _{j}^{(2)})/2$ is associated with the variation of the average energy. The second one is the tuning mode related to the energy gap between the two excited states with couplings $\kappa _{j}^{-}=(\kappa _{j}^{(1)}-\kappa _{j}^{(2)})/2$ and the third one is the inter-state coupling mode \cite{Davidson1977,MDL1985,Yarkony1998,Cederbaum2005,Gindensperger2006,Domcke2011}.

\begin{figure}
 \centering
\includegraphics[width =1.\columnwidth]{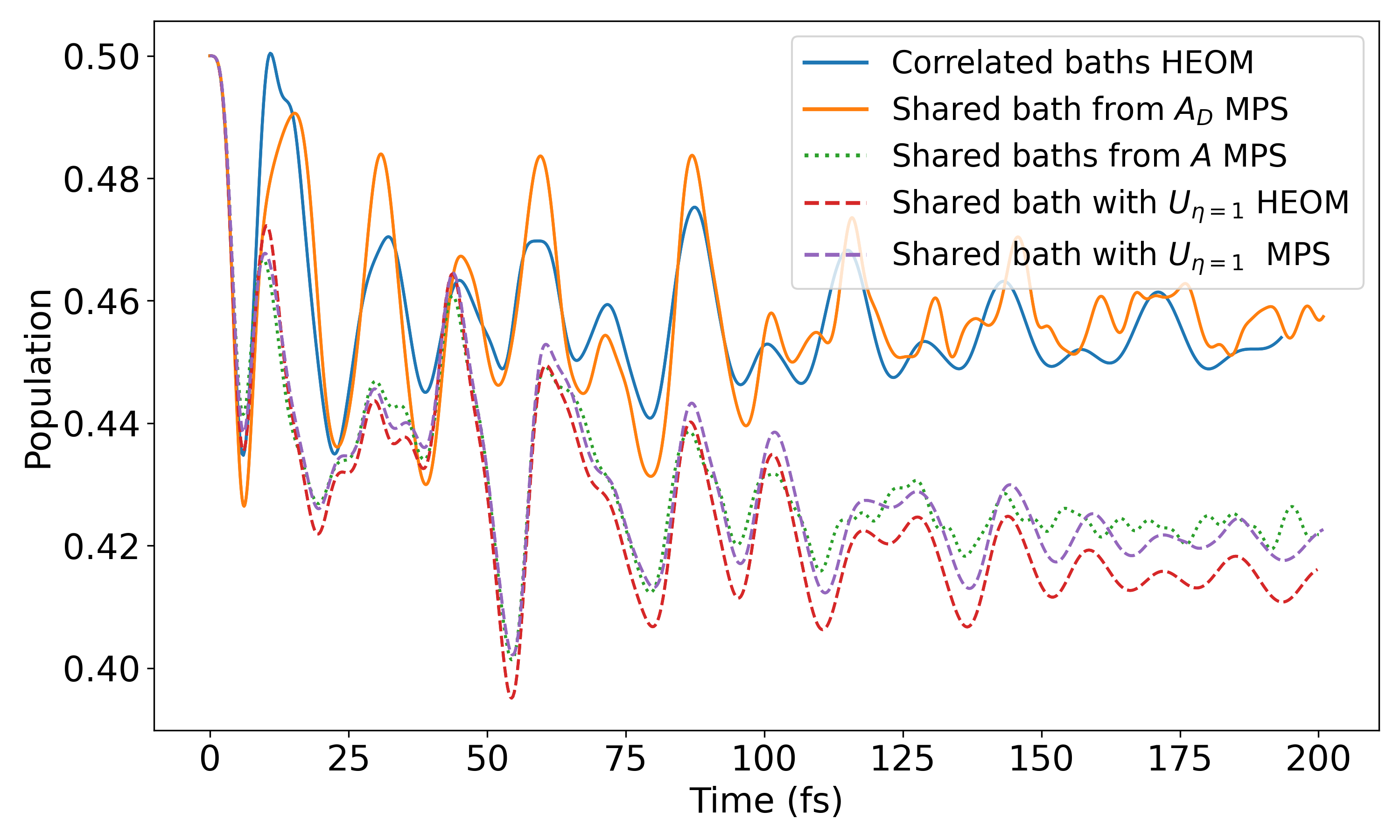}
\caption{Population $P_2$ in the $\text{S}_2$  excited state with an initial superposition with equal weights in each state. We compare the exact HEOM for correlated baths (Eq.(\ref{eq:HEOMeq})) with the T-TEDOPA simulation with a single shared bath given by the factorization $\textbf{A}_D(\omega)$ (Eq.(\ref{eq:matAD})) (full lines). The T-TEDOPA method with the two shared baths provided by the $\textbf{A}(\omega)$ (Eq.(\ref{eq:matA})) is in dots. The HEOM and MPS simulations with the approximate shared bath with $\textbf{U}_{\eta = 1}$ are in dashed lines. (The cases with $\textbf{U}_{\eta = 0.8}$ and the case discarding the correlation, i.e., with two uncorrelated tuning baths are shown in the \modif{supplementary material}.)   }  
\label{fig:popHEOMMPS}
\end{figure}

We now present the dynamics driven by the correlated baths in HEOM and by T-TEDOPA with the shared bath resulting from the $\textbf{A}_D(\omega)$ factorization. They are compared with some approximate methods. The population $P_2$ in the upper excited state is plotted in fig.\ref{fig:popHEOMMPS} for an initial superposed state with equal weights in $\text{S}_1$ and $\text{S}_2$ at $T=298~$K ($P_1=1-P_2)$.  The agreement between the correlated HEOM and the new T-TEDOPA procedure with $\textbf{A}_D(\omega)$ is very satisfactory given the absence of the Matsubara terms in HEOM. We first compare with the approximate fully correlated case, this means the shared bath generated by the $\textbf{U}_{\eta=1}$. The early dynamics is similar.  The difference from the exact result is more significant with respect to the relaxed values (the effect is still larger with $\textbf{U}_{\eta=0.8}$ given in the \modif{supplementary material}). This difference comes from a modification of the renormalization energy $\lambda_2$ when modifying the effective spectral density $J_{22}$. Indeed, in the harmonic approximation, the energy at FCP is related to the minimum energy by $\epsilon_m = \epsilon_m^0 + \lambda_m $. The variation of the difference $\lambda_1-\lambda_2$ involves a modification of $\epsilon_1^0-\epsilon_2^0$ and therefore of the equilibrated populations. Finally, we consider the T-TEDOPA simulation with the $\textbf{A}(\omega)$ factorization. As expected, the $\textbf{A}_D(\omega)$ method based on the exact diagonalization of $\textbf{J}^{\beta}(\omega)$ provides results closer to the HEOM results. The $\textbf{A}_D(\omega)$ strategy gives a behavior similar to the fully correlated case.  

\begin{figure}
 \centering
\includegraphics[width =1.\columnwidth]{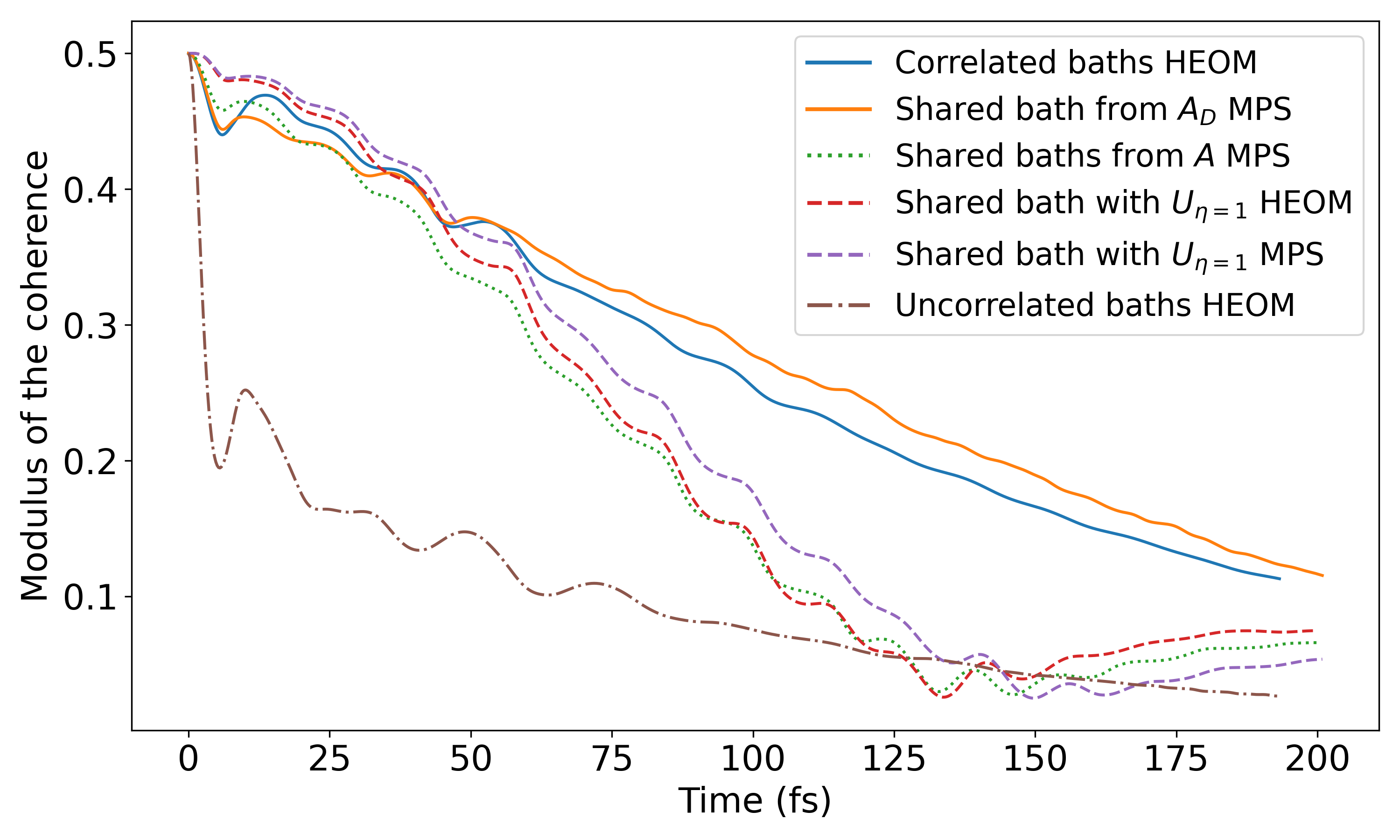}
\caption{Modulus of the electronic coherence for an initial superposition with equal weights in the $\text{S}_1$ and $\text{S}_2$ excited states. The different simulations are described in the caption of fig.\ref{fig:popHEOMMPS} and we add the uncorrelated case assuming two independent baths.
}  
\label{fig:cohe}
\end{figure}

The decay of the coherence of a superposed initial state with equal weights is shown in fig.\ref{fig:cohe}. We present the same simulations as in fig.\ref{fig:popHEOMMPS} and also the uncorrelated case that assumes two independent baths. The decoherence decay is much slower when taking into account the bath correlation. In a non-adiabatic wave packet approach, the electronic coherence is linked to the overlap of the nuclear wave packets evolving in each excited state. In the case of a sloped conical, the gradients have the same sign, the wave packet components evolve in the same direction, which promotes long-term overlap. This effect is well described by the HEOM simulation with correlated baths or by T-TEDOPA with the $\textbf{A}_D(\omega)$ formalism. The approximate fully correlated case with $\textbf{U}_{\eta = 1}$ is also acceptable in this example at least during the first fifty femtoseconds. As for the populations, the MPS method with the $\textbf{A}(\omega)$ provides results similar to the fully correlated case. On the contrary, neglecting the correlation is a very bad approximation predicting a too fast decay. 

\section{Concluding remarks}
\label{sec:conclusion}
In this work, we revisit the crossed correlation of tuning baths in a LVC model to tackle the problem of continuous spectral densities with HEOM and T-TEDOPA simulations. Indeed, in a dense vibrational environment, dynamics of the excited electronic states is driven by correlated fluctuations, both in the electrons-nuclei partition \cite{Schulten2011} and in the main vibrational coordinate mapping \cite{Egorova2012}.  

The correlated baths have been addressed early in the HEOM community \cite{Yan2002,Yan2005,SongShi2017}. In the applications on excitation energy transfer in photosynthetic aggregates, the correlation concerns baths on different sites. It is generally treated by assuming a simplified expression for the cross-correlation function \cite{Ishizaki_2005,Fleming2010,Ishizaki2010}. Here, we consider a local LVC model with a conical intersection for which an approximate correlation matrix was used in our previous works \cite{Jaouadi2022,LeDe2024,Galiana2025}. 

For the HEOM simulations, we  compare the exact correlated  propagation with an approximate method that de-correlates the baths in favorable cases by using an optimal frequency independent transformation of the tuning bath modes. The latter is chosen from an average of the eigenvectors of the $\mathbf{J}^{\beta}(\omega)$ matrix or from the eigenvectors of the initial bath correlation matrix $\mathbf{C}(t=0)$. When the dyadic structure of the discrete $\mathbf{J}^{\beta}(\omega)$ matrix is preserved during the smoothing procedure, a single eigenvector dominates that is associated with a single shared bath. This approximation is then very interesting since it requires far fewer computational resources than the correlated HEOM when the hierarchy level is high. 

Methods using the discretization of the spectral densities for MCTDH applications also face the correlation problem \cite{Burghardt2019}.  The difference with T-TEDOPA mainly lies in the sampling of the temperature dependent spectral density and in the technical way of building the chain of modes. Here, we have shown that the T-TEDOPA method allows in principle to use the exact frequency dependent transformation of the modes diagonalizing the $\mathbf{J}^{\beta}(\omega)$ matrix. However, HEOM and T-TEDOPA have very different numerical constraints. T-TEDOPA can manage the de-correlating frequency dependent transformation but the constraint lies in the length of the chains and in the long range interaction. The transformed system-bath Hamiltonian may be obtained by different factorization matrices using the uncorrelated spectral modes as suggested here or a combination of these modes as done in a recent application where the spectral densities are extracted from molecular dynamics \cite{DunnettMB2021,Hunter2024_ie,Zuehlsdorff2024}.  When the factorization matrix is not the spectral one, some dynamical correlation remains but the numerical price of the chains may be reduced. In favorable cases, T-TEDOPA using the exact spectral factorization is a very efficient alternative to the exact correlated HEOM simulation.

\section*{Supplementary Material}
The Supplementary Material gathers the parameters of the Tannor-Meier Lorentzian functions fitting the different spectral densities. We summarize the main characteristics of the bath correlation functions, specifically the case of the expansion of the complex conjugate function (Eq.(\ref{CdeTexpansion})). We also give a supplementary example of the dynamics by different approaches.    

\section{Acknowledgments}
We warmly acknowledge Dr B. Lasorne and Dr J. Galiana for the $\textit{ab}$ $\textit{initio}$ data used to calibrate the two-bath model. We also thank Dr A. Jaouadi for helpful discussions. B.LD acknowledges the support of iSiM Sorbonne. EM acknowledges support from ANR project NQESim (Grant No. ANR-23-CE29-0024-01). AWC acknowledges support from the ANR project RADPOLIMER ( ANR-22-CE30-0033 ). 

\section*{Data availability}
The data that support this study are available on request from the authors. All the T-TEDOPA simulations were done with the MPSDynamics Julia package \cite{mpsdynamics_zenodo} that can be found along with the documentation at \url{https://github.com/shareloqs/MPSDynamics}.

\bibliographystyle{unsrt}
\bibliography{bib_correl.bib}
\end{document}


\maketitle
\tableofcontents

\section{Spectral densities}
\label{sec:Specdens}

The spectral densities using the linear vibronic coupling (LVC) parameters in mass weighted coordinates are given as, 
\begin{equation}
 {{J}_{lm }}(\omega )=\frac{\pi }{2}\sum\nolimits_{k}{\frac{f_{k}^{(l)}f_k^{(m)}}{{{\omega }_{k}}}}\,\delta (\omega -{{\omega }_{k}}).
\end{equation}
In order to account for the existence of an environment, the delta distribution is broadened by a Lorentzian smoothing function,
\begin{equation}
\delta (\omega -{{\omega }_{j}})\sim\frac{1}{\pi }\frac{\Gamma}{{{\left( \omega -{{\omega }_{j}} \right)}^{2}}+{{\Gamma }^{2}}},
\label{eq:delta}
\end{equation}
leading to a continuous spectral density.
We take $\Gamma =\qty{80}{\per\centi\meter}$. 

The spectral densities of the tuning baths ($J_{11}$, $J_{22}$), their crossed term ($J_{12}$) and the coupling bath ($J_{33}$) are fitted by a sum of ${n}_\text{lor}$ Tannor-Meier Lozentzian functions,\cite{Meier1999,Pomyalov2010}
\begin{equation}
 J^\text{TM}\left( \omega  \right)  \approx\sum\limits_{l=1}^{{{n}_\text{lor}}}{\frac{{{p}_{l}}\omega }{\hbar^3\left[ {{\left( \omega +{{\Omega }_{l}} \right)}^{2}}+\Gamma _{l}^{2} \right]\left[ {{\left( \omega -{{\Omega }_{l}} \right)}^{2}}+\Gamma _{l}^{2} \right]}} \quad.
\label{eq:JTM}
\end{equation}
The parameters $p_{l}$, $\Omega_{l}$, and $\Gamma_{l}$ are given in Table \ref{tab:table_threebathsm22} for the m22 dimer. 

\begin{table}[ht]
    \begin{tabular}{cccc}
    \toprule
    
    & $p_{l}$ (Ha$^4$) & $\hbar \Omega_{l}$ (Ha) & $\hbar \Gamma_{l}$ (Ha) \\ 

\midrule
$J_{11}$ & $5.30 \times 10^{-10}$ & $1.078687 \times 10^{-2}$ & $3.4 \times 10^{-4}$ \\
~      &  $1.06 \times 10^{-10}$ & $7.6634\times 10^{-3}$ & $4.0 \times 10^{-4}$ \\
 ~      & $5.3\times 10^{-11}$ & $5.4517 \times 10^{-3}$ & $3.4 \times 10^{-4}$ \\
 $J_{22}$ & $4.3 \times 10^{-10}$ & $1.078687 \times 10^{-2}$ & $3.4 \times 10^{-4}$ \\
~      &  $1.39 \times 10^{-10}$ & $7.6634\times 10^{-3}$ & $3.8 \times 10^{-4}$ \\
 ~      & $2.9\times 10^{-10}$ & $5.4517 \times 10^{-3}$ & $8.0 \times 10^{-4}$ \\
  $J_{12}$ & $4.80 \times 10^{-10}$ & $1.078687 \times 10^{-2}$ & $3.4 \times 10^{-4}$ \\
~      &  $1.06 \times 10^{-10}$ & $7.6634\times 10^{-3}$ & $4 \times 10^{-4}$ \\
 ~      & $4.8\times 10^{-11}$ & $5.4517 \times 10^{-3}$ & $3.4 \times 10^{-4}$ \\
$J_{33}$ & $3.82 \times 10^{-10}$ & $1.078275 \times 10^{-2}$ & $3.4 \times 10^{-4}$ \\
 ~       & $1.45 \times 10^{-10}$ & $7.6634 \times 10^{-3}$ & $4.0 \times 10^{-4}$  \\
 ~       & $4.58 \times 10^{-11}$ & $5.286 \times 10^{-3}$ & $4.5 \times 10^{-4}$ \\
\bottomrule
    \end{tabular}
\caption{\label{tab:table_threebathsm22} Parameters of the Tannor-Meier Lorentzian functions fitting the spectral densities of LVC model of the m22 dimer. $J_{11}(\omega)$ and $J_{22}(\omega)$ correspond to the tuning baths that make the energies of states $S_1$ and $S_2$ fluctuate to first order (tuning). $J_{12}(\omega)$ is the crossed term. $J_{33}(\omega)$ refers to the coupling bath inducing first-order variations of the interstate coupling.\\
Alt text: A table listing the fitted parameters of the Tannor-Meier Lorentzian functions for the LVC model. The data is organized to present parameters for the tuning baths, the crossed term, and the interstate coupling bath of the dimer.}
\end{table}

The parameters of the two exponential functions
\begin{equation}
    C_{Matsubara}(t)=\sum\limits_{k=1}^{N}a_{k}\times e^{(-b_{k}*t)}.
\end{equation}
fitting the contribution of the Matsubara terms to the bath correlation functions of the m22 example at 298K are given in Table \ref{tab:table_fitting_matsu}. This procedure is suggested in ref.\cite{Lambert2019}.

\begin{table}[!ht]
    \centering
    \begin{tabular}{lccc}
    \hline
    \hline
        ~ & ~ & function 1  & function 2  
        \\ \hline
        $C^{tuning}_{Matsubara}$  & a & -3.89413e-07 & -6.22463e-07 \\ 
        ~ & b & 0.02088 & 0.006372
        \\ 
        $C^{coupling}_{Matsubara}$ & a & -4.49977e-07 & -6.76761e-07 \\ 
        ~ & b & 0.0211304 & 0.00640812       
        \\ 
        \hline
        \hline
    \end{tabular}
    \caption{\label{tab:table_fitting_matsu} Fitting parameters of the Matsubara contributions $C^{tuning}_{Matsubara}$ and $C^{coupling}_{Matsubara}$ related to the spectral densities $J_{11}$ and $J_{33}$ of the m22 model by 2 exponential functions at 298K.\\
    Alt text: A data table presenting the fitting parameters for Matsubara contributions at 298 Kelvin. Columns organize the exponential function parameters related to the tuning and coupling spectral densities of the model.}
\end{table}

\section{Correlation functions}
\label{sec:correlationfunction}
\begin{figure*}
    \centering
    \includegraphics[width=0.75\textwidth]{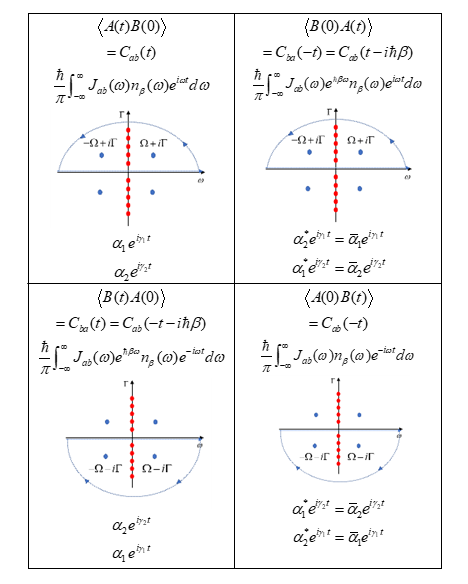}
        \caption{The four correlation functions involved in the HEOM equations with two correlated baths. The blue circles represent the four poles of a Tannor-Meier Lorentzian function fitting the spectral density (Eq(\ref{eq:TMLorentzian})). The relevant contour to compute the FT (Eq.(\ref{eq:CdetFT})) is schematized and the parameters of the expansion of the correlation function are given in terms of Eqs.(\ref{eq:alpha1} and \ref{eq:alpha2})\\
        Alt text: Four separate complex-plane diagrams, each showing an integration contour. Blue circles represent the four poles of the Lorentzian function. Red circles are the poles of the Bose function. The repeated layout illustrates the specific configuration for each of the four correlation functions.}
    \label{fig:SMCorre}
\end{figure*}

The HEOM equations with correlated baths involve the correlation between the effective modes of two baths of the type ${{C}_{ab}}(t,\tau )=\left\langle A(t)B(\tau ) \right\rangle$ and ${{C}_{ba}}(\tau ,t)=\left\langle B(\tau )A(t) \right\rangle $. We first recall the relation
\begin{align}
  & {{C}_{ba}}(\tau ,t)=\left\langle B(\tau )A(t) \right\rangle  \nonumber \\ 
 & =\left\langle A(t-i\hbar \beta )B(\tau ) \right\rangle ={{C}_{ab}}(t-i\hbar \beta ,\tau )  
\end{align}
that is known as the KMS (Kubo Martin Schwinger) theorem. Indeed, one has by the cyclic property of the trace:
\begin{align}
  & {{C}_{ba}}(\tau ,t)=\left\langle B(\tau )A(t) \right\rangle =\frac{1}{Z}Tr\left[ B(\tau ){{e}^{iHt/\hbar }}A{{e}^{-iHt/\hbar }}{{e}^{-\beta H}} \right] \nonumber \\ 
 & =\frac{1}{Z}Tr\left[ B(\tau ){{e}^{-\beta H}}{{e}^{\beta H}}{{e}^{iHt/\hbar }}A{{e}^{-iHt/\hbar }}{{e}^{-\beta H}} \right] \nonumber \\ 
 & =\frac{1}{Z}Tr\left[ {{e}^{\beta H}}{{e}^{iHt/\hbar }}A{{e}^{-iHt/\hbar }}{{e}^{-\beta H}}B(\tau ){{e}^{-\beta H}} \right] \nonumber \\ 
 & =\frac{1}{Z}Tr\left[ {{e}^{iH\left( t-i\hbar \beta  \right)/\hbar }}A{{e}^{-iH\left( t-i\hbar \beta  \right)/\hbar }}B(\tau ){{e}^{-\beta H}} \right] \nonumber \\ 
 & ={{C}_{ab}}(t-i\hbar \beta ,\tau )  
\end{align}
where $Z$ is the partition function of the canonical bath. The HEOM equations with two correlated baths involve four correlation functions, on the one hand $C_{ab}(t)$ and $C_{ba}(-t)$ and on the other hand $C_{ba}(t)$ and $C_{ab}(-t)$ (by setting $\tau =0$). Indeed, the master equation giving the reduced density matrix of the system in interaction representation is:
\begin{equation}
{{\dot{\tilde{\rho }}}_{S}}(t)=-\frac{1}{{{\hbar }^{2}}}{{\mathcal{T}}_{+}}\sum\limits_{a,b=1}^{{{N}_{bath}}}{\left[ {{S}_{a}}(t),\int\limits_{0}^{t}{d}\tau {{C}_{ab}}(t-\tau ){{S}_{b}}(\tau ){{{\tilde{\rho }}}_{S}}(t)-{{C}_{ba}}(\tau -t){{{\tilde{\rho }}}_{S}}(t){{S}_{b}}(\tau ) \right]}    \end{equation}
where ${\mathcal{T}}_{+}$ is the time ordering operator and $S_l(t)$ is the system coupling operator in interaction representation. We now illustrate how to compute the parameters of the correlation functions  ${{C}_{ab}}(t)$ and ${{C}_{ba}}(-t)$  when the spectral density is assumed to be parametrized by a single Tannor-Meier function
\begin{equation}
{{J}_{ab}}\left( \omega  \right)=\frac{p\omega }{{{\hbar }^{3}}\left[ {{\left( \omega +\Omega  \right)}^{2}}+\Gamma _{{}}^{2} \right]\left[ {{\left( \omega -\Omega  \right)}^{2}}+\Gamma _{{}}^{2} \right]}.
\label{eq:TMLorentzian}
\end{equation}
${{C}_{ab}}(t)$ is obtained by the Fourier transform (FT)
\begin{equation}
 {{C}_{ab}}(t)=\frac{\hbar }{\pi }\int_{-\infty }^{\infty }{d\omega J_{ab}^{{}}(\omega )}{{n}_{\beta }}(\omega ){{e}^{i\omega t}}
 \label{eq:CdetFT}
\end{equation}
where ${{n}_{\beta }}(\omega )={{\left( {{e}^{\hbar\beta \omega }}-1 \right)}^{-1}}$. The FT can be calculated by the residue theorem in the complex plane \cite{Pomyalov2010}. The Lorentzian function (Eq.(\ref{eq:TMLorentzian})) has four poles located at $\pm \,\Omega \pm \,i\Gamma $. The Bose function has an infinite number of poles on the imaginary axis known as the Matsubara terms. In practice, this infinite series may be truncated according to the temperature value.  As the contour must be closed in the upper or in the lower half plane, only two poles of the Lorentzian are relevant in each case. For ${C}_{ab}(t)$, the contour is closed in the upper half plane and one gets by separating the contribution of the Matsubara terms:
\begin{equation}
 {{C}_{ab}}(t)=\sum\nolimits_{k=1}^{2}{\alpha _{k}^{{}}}{{e}^{i\gamma _{k}^{{}}t}}+\sum\nolimits_{m = 1 }^{\infty }{{{\alpha }_{m}}}{{e}^{i{{\gamma }_{m}}t}}.   
\end{equation}
with
\begin{equation}
 \alpha _{1}^{{}}{{e}^{i\gamma _{1}^{{}}t}}=\frac{p}{8\hbar^2\Omega \Gamma }\left[ \coth \left( \frac{\hbar\beta }{2}(\Omega +i\Gamma ) \right)-1 \right]{{e}^{i\left( \Omega +i\Gamma  \right)t}}
 \label{eq:alpha1}
\end{equation}
and
\begin{equation}
 \alpha _{2}^{{}}{{e}^{i\gamma _{2}^{{}}t}}=\frac{p}{8\hbar^2\Omega \Gamma }\left[ \coth \left( \frac{\hbar\beta }{2}(\Omega -i\Gamma ) \right)+1 \right]{{e}^{i\left( -\Omega +i\Gamma  \right)t}}.
 \label{eq:alpha2}
\end{equation}
The Matsubara terms are given by:
\begin{equation}
 {{\gamma }_{m}}=i2\pi m/\hbar \beta    
\end{equation}
and
\begin{equation}
{{\alpha }_{m}}=2iJ({{\gamma }_{m}})/\beta.     
\end{equation}
To compute $C_{ba}(-t)$, we use the relation $C_{ba}(-t) = C_{ab}(t-i\hbar\beta)$. This means that $J_{ab}(\omega)$ is simply multiplied by ${{e}^{\hbar\beta \omega }}$ in the FT. The contour is the same, and the parameters of the two decay modes are given in fig.\ref{fig:SMCorre} as a function of Eqs.(\ref{eq:alpha1} and \ref{eq:alpha2}). This allows us to define the parameters ${{\bar{\alpha }}_{1}}$ and ${{\bar{\alpha }}_{2}}$ used to maintain the same numbering for the terms with the rate $\gamma_1$ or $\gamma_2$ respectively. 
The last two correlation functions are $C_{ba}(t)$ and $C_{ab}(-t)$. They are computed in a similar way by $C_{ba}(t)= C_{ab}(-t-i\hbar\beta)$. The contour must now be closed in the lower half plane as shown in fig.\ref{fig:SMCorre}. In summary, the contribution of a single Tannor-Meier Lorentzian function may be written as follows:
\begin{equation}
{{C}_{ab}}(t,\tau )=\sum\nolimits_{k=1}^{2}{\alpha _{k}^{(ab)}}{{e}^{i\gamma _{k}^{(ab)}(t-\tau )}}
\label{CdeTexpansion}
\end{equation}
and
\begin{align}
  & {{C}_{ba}}(\tau ,t)=C_{ab}^{*}(t,\tau ) \nonumber \\ 
 & =\sum\nolimits_{k=1}^{{2}}{\bar{\alpha }_{k}^{(lm)}}{{e}^{i\gamma _{k}^{(ab)}(t-\tau )}}  
 \label{CdeTconjg}
\end{align}
with
\begin{align}
 \bar{\alpha}_1=\alpha_2^* \nonumber \\
 \bar{\alpha}_2=\alpha_1^*
\end{align}
The contribution of the poles of the Bose function remains the same for $C_{ab}(t)$ and $C_{ba}(-t)$ so that one has
\begin{equation}
\bar{\alpha}_m = {\alpha}_m.   
\end{equation}

\section{Supplementary examples}
\begin{figure}
 \centering
\includegraphics[width =1.\columnwidth]{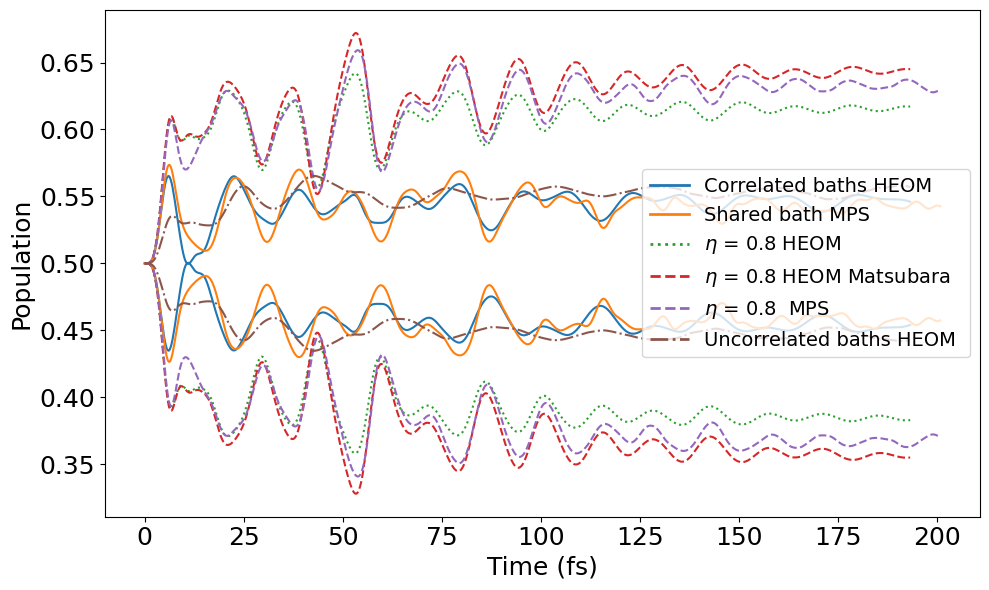}
\caption{Population in the $\text{S}_1$ and $\text{S}_2$ excited states with an initial superposed state with equal weights. We compare the correlated HEOM and T-TEDOPA simulation with the $\mathbf{A_D}$ transformation, the results with the approximation of a frequency independent de-correlation matrix $\mathbf{U}_{\eta=0.8}$ , and the case discarding the correlation, i.e., with two independent tuning baths. \\
Alt text : A line plot tracking the populations of both S1 and S2 excited states over time. Multiple overlapping curves compare different simulation methods and approximation models against the exact correlated bath case to illustrate their accuracy.  }  
\label{fig:popHEOMMPS08}
\end{figure}

Figure \ref{fig:popHEOMMPS08} compares the populations obtained by the correlated HEOM and the T-TEDOPA method using the factorization of the spectral density matrix by the $\mathbf{A}_D$ transformation with the approximate method involving the frequency independent diagonalization matrix $\mathbf{U}_{\eta=0.8}$. The value $\eta = 0.8$ ($J_{22}=\eta J_{11}$)  is chosen from the highest peaks in the spectral densities. We also present the corresponding uncorrelated case. The approximation with $\eta = 0.8$ gives similar early decay and oscillatory pattern but different asymptotic values. This is linked to the approximation of the $J_{22}$ spectral density inducing a different renormalization energy. In this case, with the approximation of a frequency independent transformation $\mathbf{U}_{\eta=0.8}$ the T-TEDOPA and the HEOM simulation are in good agreement when the Matsubara terms are included.  The uncorrelated bath assumption has little impact on the population but, as shown in the main text, the decay of the coherence is completely different. 

%

\clearpage
\bibliography{bib_correl.bib}